\documentclass[prd,nofootinbib,aps,twocolumn,showpacs,preprintnumbers,amsmath,amssymb,floatfix]{revtex4-1}
\usepackage{graphicx}
\usepackage{xspace}
\usepackage{booktabs}
\usepackage{epstopdf}
\usepackage{color}
\newcommand{\be}{\begin{equation}}
\newcommand{\ee}{\end{equation}}
\newcommand{\benn}{\begin{equation*}}
\newcommand{\eenn}{\end{equation*}}
\newcommand{\bea}{\begin{equation}\begin{aligned}}
\newcommand{\eea}{\end{aligned}\end{equation}}
\newcommand{\beann}{\begin{equation*}\begin{aligned}}
\newcommand{\eeann}{\end{aligned}\end{equation*}}
\newcommand{\bse}{\begin{subequations}}
\newcommand{\ese}{\end{subequations}}
\newcommand{\Order}{{\cal O}}   
\newcommand{\ie}{i.e.}
\newcommand{\SM}{SM}
\newcommand{\PQ}{PQ}
\newcommand{\eq}{\ensuremath{{\text{eq}}}}
\newcommand{\Reheating}{\mathrm{R}}
\newcommand{\TP}{\mathrm{TP}}
\newcommand{\NTP}{\mathrm{NTP}}
\newcommand{\vip}{\tilde{v}_{i}}
\newcommand{\seconds}{\mathrm{s}}
\newcommand{\eV}{\mathrm{eV}}
\newcommand{\meV}{\mathrm{meV}}
\newcommand{\keV}{\mathrm{keV}}
\newcommand{\MeV}{\mathrm{MeV}}
\newcommand{\GeV}{\mathrm{GeV}}
\newcommand{\TeV}{\mathrm{TeV}}
\newcommand{\axion}{\ensuremath{a}}
\newcommand{\saxion}{\ensuremath{\sigma}\xspace}
\newcommand{\axino}{\ensuremath{\tilde{a}}\xspace}
\newcommand{\ax}{\ensuremath{_{a}}}
\newcommand{\sax}{\ensuremath{_{\saxion}}}
\newcommand{\electron}{\ensuremath{e^-}}
\newcommand{\maxion}{\ensuremath{m_{a}}}
\newcommand{\msaxion}{\ensuremath{m_\saxion}\xspace}
\newcommand{\maxino}{\ensuremath{m_{\tilde{a}}}\xspace}
\newcommand{\mgravitino}{\ensuremath{m_{3/2}}\xspace}
\newcommand{\tausaxion}{\ensuremath{\tau_\saxion}\xspace}
\newcommand{\TNR}{T_{\mathrm{nr}}}
\newcommand{\TD}{T_{\mathrm{D}}}
\newcommand{\TR}{T_{\Reheating}}
\newcommand{\tr}{\ensuremath{T_{\text{R}}}}

\newcommand{\fax}{\ensuremath{f_{\mathrm{PQ}}}}
\newcommand{\vax}{\ensuremath{v_{\mathrm{PQ}}}}
\newcommand{\fb}{\ensuremath{f_{\text{B}}}}

\newcommand{\kcut}{\ensuremath{k_{\text{cut}}}}
\newcommand{\YaxTP}{Y_{\axion}^{\mathrm{TP}}}
\newcommand{\Yaxeq}{Y_{\axion}^{\mathrm{eq}}}
\newcommand{\YaxeqTP}{Y_{\axion}^{\mathrm{eq/TP}}}
\newcommand{\YsaxTP}{Y_{\saxion}^{\mathrm{TP}}}
\newcommand{\YsaxeqTP}{Y_{\saxion}^{\mathrm{eq/TP}}}
\newcommand{\OmegaDM}{\Omega_{\mathrm{CDM}}}
\newcommand{\omegaDM}{\omega_{\mathrm{CDM}}}
\newcommand{\nax}{n_{\axion}}
\newcommand{\nsax}{n_{\saxion}}

\newcommand{\neff}{\ensuremath{N_\text{eff}}}
\begin{document}
%
%
\preprint{arXiv:1208.2951}
\preprint{MPP-2012-104}
%
%
%
\title{Axions and saxions from the primordial supersymmetric plasma\\
and extra radiation signatures}
\author{Peter Graf}
\affiliation{Max-Planck-Institut f\"ur Physik, 
F\"ohringer Ring 6,
D--80805 Munich, Germany}
\author{Frank Daniel Steffen}
\affiliation{Max-Planck-Institut f\"ur Physik, 
F\"ohringer Ring 6,
D--80805 Munich, Germany}
%
%
\begin{abstract}
  We calculate the rate for thermal production of axions and saxions 
  via scattering of quarks, gluons, squarks, and gluinos 
  in the primordial supersymmetric plasma.
  Systematic field theoretical methods such as hard thermal loop
  resummation are applied
  to obtain a finite result in a gauge-invariant way that is
  consistent to leading order in the strong gauge coupling.
  We calculate the thermally produced yield 
  and the decoupling temperature for both axions and saxions. 
  For the generic case in which saxion decays into axions are possible,
  the emitted axions can constitute extra radiation 
  already prior to big bang nucleosynthesis and well thereafter.
  We update associated limits imposed by recent studies 
  of the primordial helium-4 abundance
  and by precision cosmology 
  of the cosmic microwave background and large scale structure. 
  We show that the trend towards extra radiation seen in those studies
  can be explained by late decays of thermal saxions into axions
  and that upcoming Planck results will probe supersymmetric axion models 
  with unprecedented sensitivity.
\end{abstract}
%
%
\maketitle
%
%
\section{Introduction}
There are several hints towards physics beyond the standard model (\SM). 
One of them is the strong CP problem. 
If this problem is solved via the Peccei--Quinn (\PQ) mechanism, 
the axion $\axion$ arises 
as the pseudo-Nambu-Goldstone boson associated
with the U(1)$_\mathrm{PQ}$ symmetry 
broken spontaneously at the \PQ\ scale $\fax$~\cite{Sikivie:2006ni,Kim:2008hd}.
Another attractive extension of the \SM\ is supersymmetry 
(SUSY)~\cite{Martin:1997ns,Drees:2004jm,Baer:2006rs,Dreiner:2008tw}.
In conceivable settings with both the \PQ\ mechanism and SUSY,
the pseudo-scalar axion is part of a supermultiplet in which also 
its scalar partner, the saxion $\saxion$, 
and its fermionic partner, the axino $\axino$, appear.
The energy density of the early Universe can then receive contributions
from coherent oscillations 
of the axion field~\cite{Beltran:2006sq,Sikivie:2006ni,Kim:2008hd} 
and the saxion field~\cite{Chang:1996ih,Hashimoto:1998ua,Asaka:1998ns,Kawasaki:2007mk,Kawasaki:2011aa}
and from thermal production of 
axions~\cite{Turner:1986tb,Chang:1993gm,Masso:2002np,Hannestad:2005df,Sikivie:2006ni,Graf:2010tv}, saxions~\cite{Kim:1992eu,Chang:1996ih,Asaka:1998ns}, and 
axinos~\cite{Rajagopal:1990yx,Bonometto:1993fx,Chun:1995hc,Asaka:2000ew,Covi:2001nw,Brandenburg:2004du,Strumia:2010aa,Chun:2011zd,Bae:2011jb,Choi:2011yf,Bae:2011iw}
in the hot primordial plasma.

Here we calculate for the first time the thermal production rate 
of axions and saxions
via scattering processes of quarks, gluons, squarks, and gluinos
in a gauge-invariant way consistent to leading order in the strong coupling constant $g_s$.
In our calculation we use hard thermal loop (HTL) resummation~\cite{Braaten:1989mz} 
and the Braaten--Yuan prescription~\cite{Braaten:1991dd} 
to account systematically for screening effects in the quark-gluon-squark-gluino plasma (QGSGP). 
This method was introduced on the example of axion production 
in a hot QED plasma~\cite{Braaten:1991dd}; 
see also~Ref.~\cite{Bolz:2000fu}.
Moreover, it has been applied to calculate the thermal production of 
gravitinos~\cite{Bolz:2000fu,Pradler:2006qh,Pradler:2006hh,Pradler:thesis} 
and axinos~\cite{Brandenburg:2004du} in SUSY settings 
and of axions in a non-SUSY quark-gluon plasma 
(QGP)~\cite{Graf:thesis,Graf:2010tv}.

Based on our result for the thermal axion/saxion production rate, 
we determine the respective thermally produced yields and
estimate the decoupling temperature of axions and saxions 
from the thermal bath. 
While both axions and axinos are promising dark matter candidates
(cf.~\cite{Sikivie:2006ni,Kim:2008hd,Steffen:2008qp,Covi:2009pq}
and references therein),
saxions can be late decaying particles with potentially 
severe cosmological implications.
For example, energetic hadrons and photons from saxion decays 
during or after big bang nucleosynthesis (BBN) can change the abundances 
of the primordial light elements~\cite{Kawasaki:2007mk}.
Moreover, photons from saxion decays can affect the black body spectrum 
of the cosmic microwave background (CMB) for a saxion lifetime of 
$10^6\,\seconds\lesssim \tausaxion\lesssim 10^{13}\,\seconds$~\cite{Asaka:1998xa,Kawasaki:2007mk}
or may contribute either to the diffuse $X(\gamma)$-ray background 
or as an additional source of reionization for
$\tausaxion\gtrsim 10^{13}\,\seconds$~\cite{Kawasaki:1997ah,Asaka:1998xa,Chen:2003gz,Kawasaki:2007mk}.
In scenarios in which the decay mode into axions is not the dominant one,
saxion decays may also produce significant amounts of 
entropy~\cite{Kim:1992eu,Lyth:1993zw,Hashimoto:1998ua,
Asaka:1998xa,Hasenkamp:2010if,Baer:2010gr}.
This can dilute relic densities of species decoupled from the plasma and also
the baryon asymmetry $\eta$. Then, $\tausaxion<1~\seconds$ is imposed
by successful BBN which requires a standard thermal history 
for temperatures below $T\sim1~\MeV$.

In this work, however, we look at scenarios 
in which saxions (from thermal processes)
decay predominantly into axions.
Moreover, we still focus on decays prior to BBN 
and compute the additional radiation 
provided in the form of the emitted relativistic axions.
Such a non-standard contribution $\Delta\neff$
to the effective number of light neutrino species $\neff$
from decays of thermal saxions into axions  
was previously considered
in Refs.~\cite{Chun:1995hc,Chang:1996ih,Choi:1996vz,Kawasaki:2007mk}.
Applying our new result for the thermally produced saxion yield
and new cosmological constraints on $\Delta\neff$
imposed by recent studies of BBN, the CMB, and large scale 
structure (LSS)~\cite{Izotov:2010ca,Aver:2010wq,Hamann:2010pw},
we present updated 
limits on the \PQ\ scale $\fax$, the saxion mass $\msaxion$, 
and the reheating temperature $\TR$ after inflation.

Interestingly, precision 
cosmology~\cite{Hamann:2007pi,Reid:2009nq,Komatsu:2010fb,Hamann:2010pw,GonzalezGarcia:2010un} 
and recent studies of the primordial $^4$He abundance~\cite{Izotov:2010ca,Aver:2010wq} 
show a trend towards a radiation content that exceeds the predictions of the SM. 
In fact, such an excess can be explained by the considered saxion decays into axions.
The observed trend may thus be a hint for the existence of a SUSY axion model.
Here results from the Planck satellite mission will be extremely valuable, 
which will come with an unprecedented sensitivity to the amount of extra radiation at times 
much later than those at which BBN probes this quantity.
Based on a forecasted 68\% confidence level (CL) sensitivity of 
$\Delta\neff=0.26$~\cite{Perotto:2006rj,Hamann:2007sb}, 
we indicate parameter regions of SUSY axion models 
that will be tested by results from the Planck satellite mission
 expected to be published in the near future.

The remainder of this paper is organized as follows. 
In the next section we consider interactions of the PQ supermultiplet 
and decay widths for saxion decays.
In Sects.~\ref{Sec:ThermalSaxionProduction} and~\ref{Sec:ThermalAxionProduction}
our calculations of the thermal production rates of saxions and axions are presented.
We compute the associated yields in Sect.~\ref{Eq:SaxionAxionYield} 
and use the results to estimate the saxion/axion decoupling temperature in Sect.~\ref{Sec:TD}.
Then we explore $\Delta\neff$ provided in the form of axions from saxion decays
and possible manifestations in studies of BBN and of the CMB and LSS.
Here we comment on potential restrictions 
which can emerge from overly efficient thermal gravitino/axino production
and describe exemplary settings 
that allow for a high reheating temperature of 
$\TR\sim 10^8$--$10^{10}\,\GeV$.
In Sect.~\ref{Sec:AxionDensity} we compare the relic density 
of axions from the misalignment mechanism 
with the ones of thermal axions
and of non-thermal axions from saxion decays.
Our conclusions are given in Sect.~\ref{Sec:Conclusion}.

\section{Particle physics setting}
\label{Sec:Setting}
%
In a SUSY framework, the U(1)$_\mathrm{PQ}$ symmetry 
is extended to a symmetry of the (holomorphic) superpotential 
and thereby to its complex form U(1)$_\mathrm{PQ}^c$~\cite{Kugo:1983ma}.
In the case of unbroken SUSY, this implies the existence of a flat direction
and thereby a massless saxion field.
Once SUSY is broken, this flat direction gets lifted, 
resulting in a model-dependent mass of the saxion \msaxion.
For example, \msaxion is expected to be of the order of the gravitino mass \mgravitino
in gravity-mediated SUSY breaking models~\cite{Kim:1992eu,Asaka:1998ns,Asaka:1998xa}. 
Here we do not look at a specific model but treat \msaxion as a free parameter.

In this work we consider the particle content of the minimal supersymmetric SM (MSSM) 
extended by the PQ superfield
$A=(\saxion+ia)/\sqrt{2}+\sqrt{2}\theta\axino+F_A \theta\theta$, 
where $\theta$ denotes the corresponding fermionic superspace coordinate 
and $F_A$ the chiral auxiliary field. 
The interactions of $A$ with the color-field-strength superfield
$W^b
=
\tilde{g}^b
+D^b\theta-\sigma^{\mu\nu}\theta G^b_{\mu\nu}
+i\theta\theta\sigma^\mu D_\mu^{bd}\bar{\tilde{g}}^d$
are given by the effective Lagrangian
\be
\mathcal{L}_\text{\PQ}^\text{int} 
= 
-\frac{\sqrt{2}\alpha_s}{8\pi\fax}\int d^2\theta A\,W^bW^b + \text{h.c.}
\label{Eq:AWW}
\ , 
\ee
where $\tilde{g}^b$ is the gluino field, 
$D^b$ the real color-gauge auxiliary field,
${G}^b_{\mu\nu}$ the gluon-field-strength tensor,  
$D_\mu^{bd}=\partial_{\mu}\delta^{bd}-g_{s}f^{bcd}G_{\mu}^{c}$ 
the corresponding color-gauge covariant derivative
with color indices  $b$, $c$, and $d$,
the SU(3)$_c$ structure constants $f^{bcd}$,
and the gluon field $G_{\mu}^{c}$,
and $\alpha_s=g_s^2/(4\pi)$.
After performing the integration, 
we get for the propagating fields in 4-component spinor notation:
\begin{align}
\mathcal{L}_\text{\PQ}^\text{int} 
= 
&\frac{\alpha_s}{8\pi\fax} 
\bigg[ 
\saxion 
\left( G^{b\, \mu\nu}G^b_{\mu\nu} 
- 2D^b D^b
- 2i\bar{\tilde{g}}^b_M\gamma^\mu D_\mu^{bd}\tilde{g}_M^d \right) 
\nonumber\\
&\quad\,\,\,\,
+a
\left( G^{b\,\mu\nu}\widetilde{G}^b_{\mu\nu} 
+ 2\bar{\tilde{g}}_M^b\gamma^\mu\gamma^5D_\mu^{bd}\tilde{g}_M^d \right) 
\nonumber\\ 
&\quad\,\,\,\,
- i\bar{\axino}_M \frac{[\gamma^\mu,\gamma^\nu]}{2}\gamma^5\tilde{g}_M^b G^b_{\mu\nu} 
+2\bar{\axino}_M D^b\tilde{g}_M^b \bigg]
\ ,
\label{Eq:eff_lag}
\end{align}
where $\widetilde{G}^b_{\mu\nu}=\epsilon_{\mu\nu\rho\sigma}G^{b\,\rho\sigma}/2$
and $D^b=-g_s \sum_{\tilde{q}} \tilde{q}_i^* T_{ij}^b \tilde{q}_j$
with a sum over all squark fields $\tilde{q}$ and the SU(3)$_c$ generators $T_{ij}^b$
in their fundamental representation;
the subscript $M$ indicates 4-component Majorana spinors. 
Note that we use the space-time metric $g_{\mu\nu}=g^{\mu\nu}=\mathrm{diag}(+1,-1,-1,-1)$ 
and other conventions and notations of Ref.~\cite{Dreiner:2008tw}
and -- except for a different sign of the Levi-Civita tensor 
$\epsilon^{0123}=+1$ -- of Ref.~\cite{Drees:2004jm}.
To stress the absence of a quartic axion-gluon-gluino-gluino vertex and
for comparisons with similar $\mathcal{L}_\text{\PQ}^\text{int}$ expressions 
given in Refs.~\cite{Strumia:2010aa,Choi:2011yf},
we remark that the second term in the brackets in the second line of~\eqref{Eq:eff_lag}
can be written as 
$2\bar{\tilde{g}}_M^b\gamma^\mu\gamma^5 \partial_\mu\tilde{g}_M^b
=\partial_\mu(\bar{\tilde{g}}_M^b\gamma^\mu\gamma^5\tilde{g}_M^b)
=D_\mu^{bd}(\bar{\tilde{g}}_M^b\gamma^\mu\gamma^5\tilde{g}_M^d)$.
However, our result for the saxion-gluino-interaction term differs from the 
corresponding terms in~\cite{Strumia:2010aa} and~\cite{Choi:2011yf} 
by factors of $-2$ and $-1$, respectively.
Moreover, our findings for the axino interactions differ by a factor of $-1$ 
from the ones in~\cite{Strumia:2010aa,Choi:2011yf}.
This may result partially from metric conventions:
If we translate~\eqref{Eq:eff_lag} into the corresponding expression valid for
$g_{\mu\nu}=g^{\mu\nu}=\mathrm{diag}(-1,+1,+1,+1)$
using Appendix~A of Ref.~\cite{Dreiner:2008tw},
the sign of our result for the axino-gluino-gluon-interaction term will change,
whereas all other terms in~\eqref{Eq:eff_lag} will not be affected.

In the following we focus on hadronic 
or KSVZ axion models~\cite{Kim:1979if,Shifman:1979if} 
in a SUSY setting in which the effective Lagrangian~\eqref{Eq:eff_lag} describes 
the relevant interactions even in a conceivable very hot early stage 
of the primordial plasma with temperatures $T$ not too far below $\fax$. 
Note that we do not consider scenarios 
with a radiation-dominated epoch with $T$ 
above the masses of the heavy KSVZ (s)quarks $m_{Q,\tilde{Q}}$ 
such as those considered in Ref.~\cite{Bae:2011jb}. 

Next we address interactions between axions and saxions
in models with $N$ SM-gauge singlet \PQ\ multiplets $\Phi_i$
with \PQ\ charges $q_i$ and 
vacuum expectation values (VEVs) $\langle\phi_{i}\rangle=v_i$ 
that break the \PQ\ symmetry. 
This breaking leads to $N-1$ combinations of the multiplets $\Phi_i$
with large masses of $\Order(\vax)$ and one combination that gives
the light axion multiplet 
$\Phi=\sum_{i}q_{i}v_{i}\Phi_{i}/\vax$,
where $\vax=\sqrt{\sum_i v^2_iq_i^2}$ results 
from the requirement of canonically normalized kinetic terms 
for the axion and the saxion; cf.\ \eqref{Eq:axion_saxion} below.
To describe processes at energy scales well below $\vax$,
the heavy combinations can be integrated out  and 
the scalar parts of $\Phi_i$ can be parametrized near the VEVs as
\be
\phi_i 
= 
v_i\exp\left[\frac{q_i(\saxion + i\axion)}{\sqrt{2}\vax}\right]
\label{Eq:phiParametrization}
\ .
\ee
Here the canonical \PQ\ charge normalization requires $q_i^2=1$ 
for the smallest $q_i$.
From the kinetic terms of the \PQ\ fields,
one then determines $\vax$ as given above and
finds that interactions between axions and saxions 
can emerge as follows~\cite{Chun:1995hc}
\begin{align}
\mathcal{L}_\text{\PQ}^\text{kin} 
=& 
\sum_{i=1}^N 
\partial^\mu\phi_i\partial_\mu\phi^*_i 
\nonumber \\
\sim& 
\left( 1+\frac{\sqrt{2}x}{\vax}\saxion \right)
\left[ 
\frac{1}{2}\partial^{\mu} a\partial_\mu a
+\frac{1}{2}\partial^{\mu} \saxion\partial_\mu \saxion 
\right] 
+ \dots
\label{Eq:axion_saxion}
\end{align}
where $x=\sum_i q_i^3v_i^2/\vax^2$. 
The strength of these interactions thus depends on the model.
For example, 
$x=(v_1^2-v_2^2)/\vax^2$
in models whose superpotentials contain the term
$\kappa R (\Phi_1\Phi_2-\vax^2/2)$ 
with a Yukawa coupling $\kappa$, two \PQ\ fields with $q_1=-q_2=1$
and a SM-gauge singlet field $R$ with $q_R=0$. 
This illustrates that $x\ll 1$ is possible
if $v_1\simeq v_2\simeq\vax/\sqrt{2}$~\cite{Chun:1995hc,Kawasaki:2007mk,Kawasaki:2011ym}.
On the other hand, in a KSVZ axion model with just one \PQ\ scalar 
(with $v=\vax$ and $q=1$)~\cite{Asaka:1998ns}, one finds $x=1$, 
which is the value that we will consider in Sect.~\ref{Sec:AddRad} below.

Let us now relate the scale $\vax$, imposed by canonically normalized kinetic terms,
to $\fax$, defined by the form of the prefactor 
of the effective axion-gluon interactions in~\eqref{Eq:eff_lag}.
In a KSVZ model, those interactions emerge from axion couplings
to the heavy KSVZ quarks which are described by contributions to the superpotential
of the form 
$h\Phi_1 Q_L \bar{Q}_R$ 
with a Yukawa coupling $h$, $q_1=1$ 
and heavy quark multiplets $Q_L$ and $\bar{Q}_R$ 
with color charge and \PQ\ charges $q_Q=-1/2$.
Considering the resulting Lagrangian that describes 
the interactions of $\phi_1$ with the KSVZ quarks,
one sees that the spontaneous breaking of the \PQ\ symmetry 
results in heavy Dirac KSVZ quarks with a mass
$m_{Q} = h v_1$.
Integrating out loops of such heavy quarks, 
one finds the effective Lagrangian describing axion interactions with gluons
\be
\mathcal{L}^\text{int} 
= 
\frac{hv_1\alpha_s}{8\pi\sqrt{2} \, m_Q\vax}\,
a \, G^{b\,\mu\nu}\widetilde{G}^b_{\mu\nu} ,
\label{Eq:AxionGluon}
\ee
where the KSVZ quarks have been assumed to be 
in the fundamental representation of SU(3)$_c$.
For $\fax =\sqrt{2} \vax$,
one thus recovers the well-known form of the corresponding interaction term 
as given in~\eqref{Eq:eff_lag}.%
\footnote{Here we focus on $N_Q=1$ 
heavy KSVZ (s)quark multiplets $Q_L$ and $\bar{Q}_R$.
For $N_Q>1$, $\fax\to\fax/N_Q$, 
e.g., in~\eqref{Eq:AWW}, \eqref{Eq:eff_lag},
\eqref{Eq:fa_Limit}, and~\eqref{Eq:SaxionGluGlu} 
in line with an additional factor of
$N_Q$ on the right-hand side of~\eqref{Eq:AxionGluon}. 
Using this $\fax$ definition, there are no modifications of 
the relation $\fax =\sqrt{2}\vax$ 
and of~\eqref{Eq:sax_lifetime} below for $N_Q>1$.}

Note that an alternative convention with
$\langle\phi_{i}\rangle=\vip/\sqrt{2}$ 
and $\fax=\sqrt{\sum_i \vip^{2} q_i^2}$
can be found in the literature~\cite{Chun:2000jr}.
Then, $\phi_{i}=(\vip/\sqrt{2})\exp[q_i(\sigma+ia)/\fax]$.
Indeed, with this convention, one arrives directly
at an agreement of~\eqref{Eq:AxionGluon} with
the corresponding term in~\eqref{Eq:eff_lag}.
However, we prefer to work explicitly with both
$\fax$ and $\vax$ also to allow for a direct comparison
with literature that uses the
parametrization given in~\eqref{Eq:phiParametrization}
or a directly related one;
see e.g.\ Refs.~\cite{Asaka:1998ns,Asaka:1998xa} 
or~\cite{Ichikawa:2007jv,Kawasaki:2007mk,Moroi:2012vu}
in which their $\fax$ or $F_a$ agree with our $\vax$.

Numerous laboratory, astrophysical, and cosmological studies point 
to~\cite{Raffelt:2006cw,Beringer}
\be
\fax \gtrsim 6 \times 10^8~\GeV .
\label{Eq:fa_Limit}
\ee
This corresponds to an upper limit of about $10~\meV$ 
on the axion mass,
\be
\maxion
\simeq 
6~\meV 
\left(\frac{10^{9}\,\GeV}{\fax}\right),
\label{Eq:maxion}
\ee
and implies that axions 
are stable on cosmological timescales.
Because of the larger mass of the saxion, its lifetime $\tausaxion$ is 
typically smaller than the age of the Universe 
and governed by the following decay widths.
From~\eqref{Eq:axion_saxion} one obtains the width for the saxion decay into axions,%
\footnote{Our result~\eqref{Eq:sax_lifetime} agrees with 
the ones of Refs.~\cite{Asaka:1998ns,Asaka:1998xa}, where $x=1$ and $\fax\equiv\vax$,
and of Refs.~\cite{Ichikawa:2007jv,Kawasaki:2007mk,Moroi:2012vu}, where $F_a\equiv\vax$.}
\be
\Gamma_{\saxion\to aa} 
= \frac{x^2\msaxion^3}{64\pi \vax^2}
= \frac{x^2 \msaxion^3}{32\pi \fax^2},
\label{Eq:sax_lifetime}
\ee
and from~\eqref{Eq:eff_lag} the width for the saxion decay into gluons,
\be
\Gamma_{\saxion\to gg} = \frac{\alpha_s^2\msaxion^3}{16\pi^3\fax^2}.
\label{Eq:SaxionGluGlu}
\ee
For KSVZ fields that carry an non-zero electrical charge $e_Qe$ 
with $e=\sqrt{4\pi\alpha}$ 
and the fine-structure constant $\alpha$,
the saxion can decay into photons via KSVZ quark loops. 
After integrating out those loops, we find the associated width
\be
\Gamma_{\saxion\to\gamma\gamma} 
= 
\frac{9 e_Q^4\alpha^2 \msaxion^3}{64\pi^3\fax^2}.
\label{Eq:SaxionPhoPho}
\ee

If $x\gtrsim 0.2$, the saxion decay into axions governs $\tausaxion$,
which is the case on which we focus in this work.
Indeed, in the region with $\msaxion \gtrsim 10~\GeV$
in which the competing decay $\saxion\to gg$ is possible,
such $x$ values imply the branching ratio
$\mathrm{BR}(\saxion\to aa)\gtrsim 0.9$.
For $\msaxion$ below the threshold to form hadrons,
where $\saxion\to \gamma\gamma$ is the competing decay, 
the decay into axions governs $\tausaxion$ 
for even smaller values of $x$,
e.g., for $e_Q=1$ and $x=0.02$, 
we still find the branching ratio 
$\mathrm{BR}(\saxion\to aa)\gtrsim 0.9$.

\section{Thermal saxion production}
\label{Sec:ThermalSaxionProduction}

Let us now calculate the thermal production of saxions 
in the primordial SUSY QCD plasma.
Assuming that inflation has governed the earliest moments of the Universe,
any initial population of saxions has been diluted away by the exponential 
expansion during the slow-roll phase. 
After completion of the reheating phase 
that leads to a radiation-dominated epoch
with an initial temperature $\TR$, 
the thermal production of saxions starts to become efficient.
In fact, we focus on cosmological settings 
in which radiation governs the energy density of the Universe 
as long as this production mechanism is efficient 
(\ie, for $T$ down to at least $T\sim 0.01~\TR$).
While inflation models can point to $\TR$ well above $10^{10}~\GeV$,
we consider the case $\TR<\fax$ such that no \PQ\ symmetry restoration
takes place after inflation.
Moreover, $\TR<m_{Q,\tilde{Q}}$ is assumed in line with our comments
on the considered KSVZ axion model settings in the previous section.

\begin{figure}[h!]
\begin{description}
\item [Process A] $g^a+ g^b \rightarrow g^c + \saxion$
\begin{center}
\includegraphics[width=0.28\textwidth]{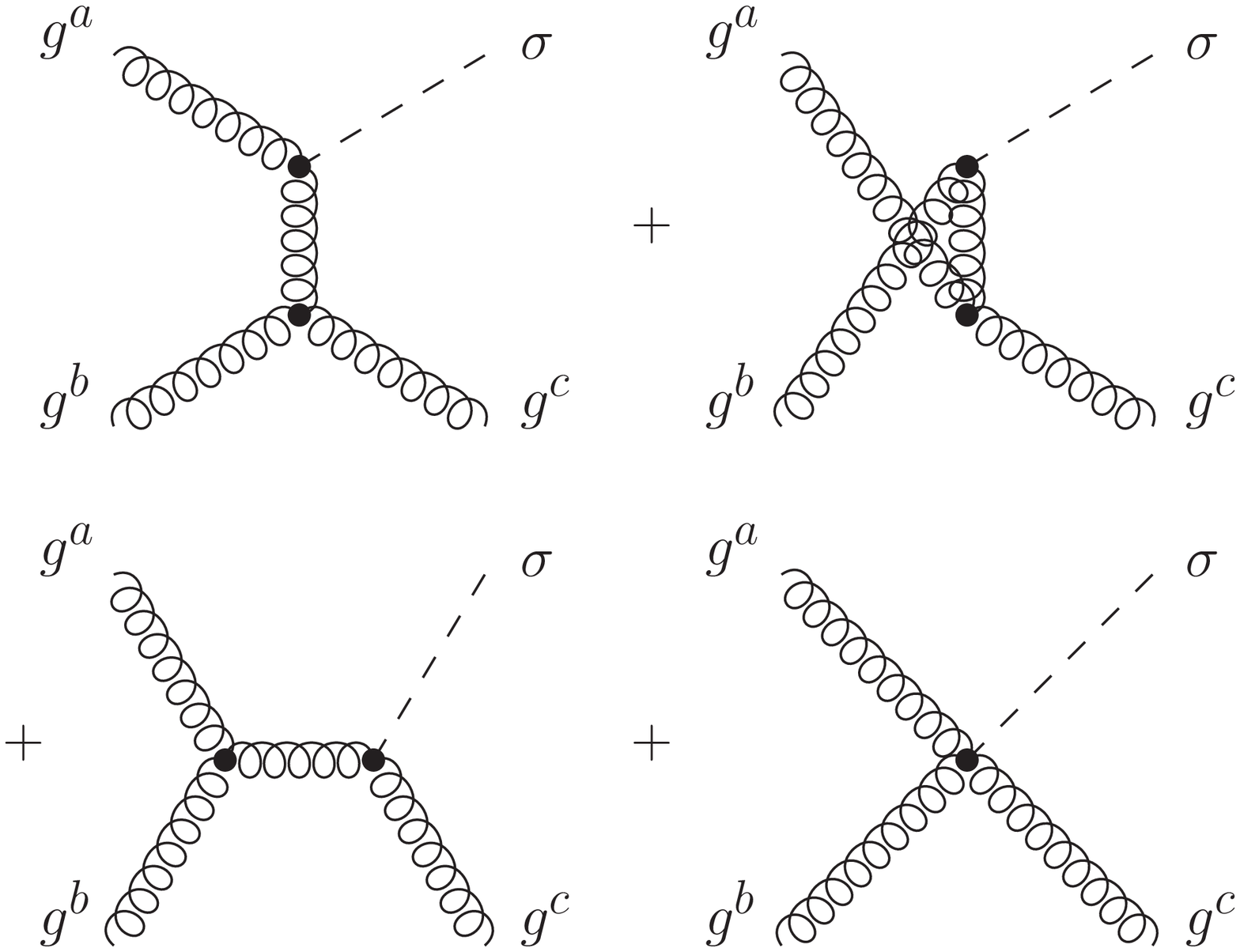}
\end{center}
\item [Process B] $q_i + \bar{q}_j \rightarrow g^a + \saxion$
\begin{center}
\includegraphics[width=0.13\textwidth]{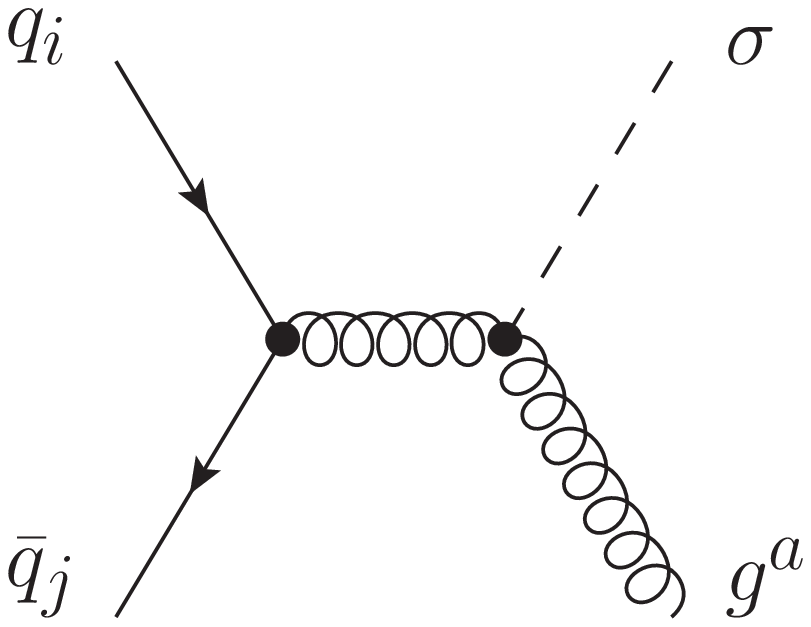}
\end{center}
\item [Process C] $q_i + g^a \rightarrow q_j + \saxion$ (Crossing of B)
\item [Process D] $\tilde{g}^a + \tilde{g}^b \rightarrow g^c + \saxion$
\begin{center}
\includegraphics[width=0.28\textwidth]{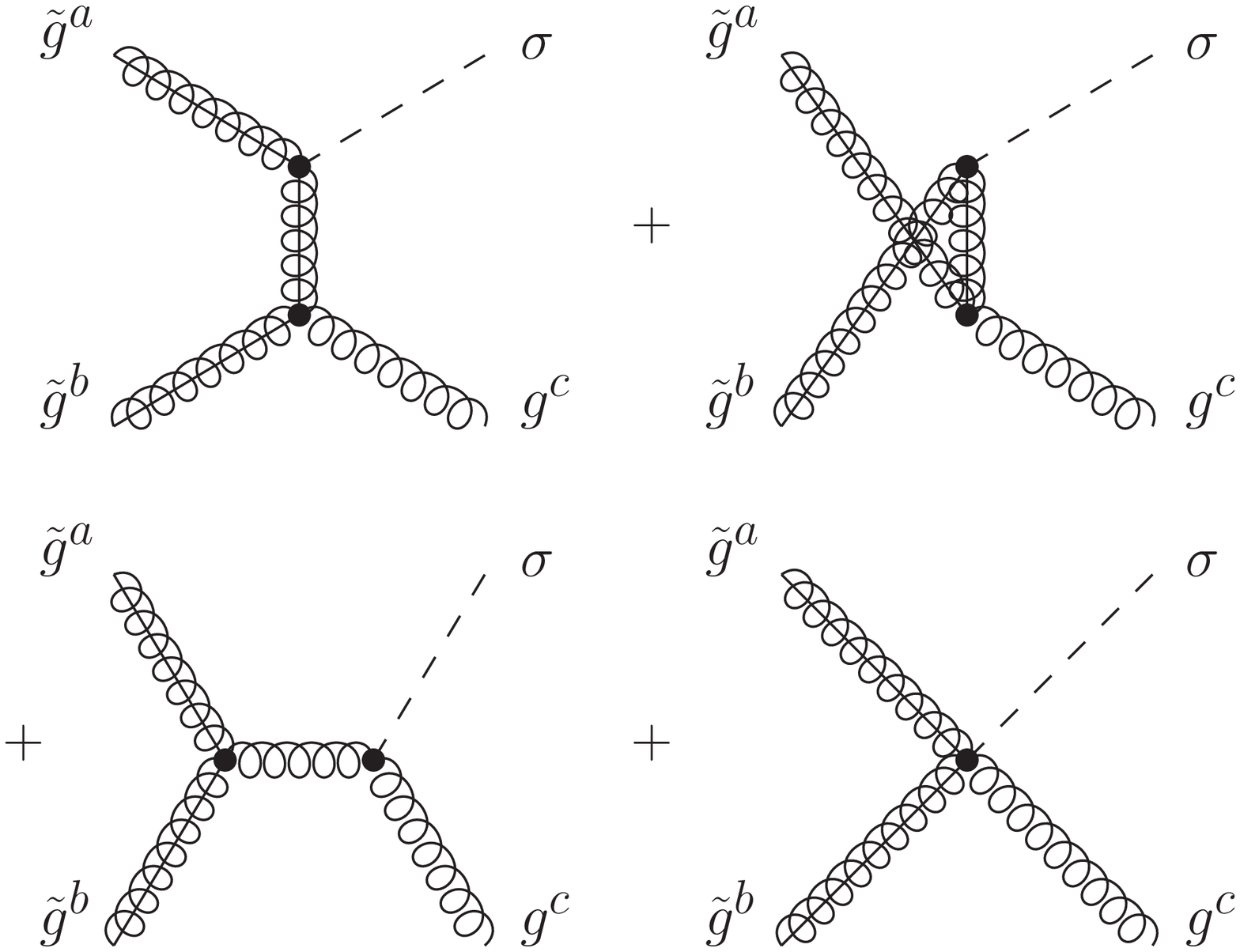}
\end{center}
\item [Process E] $\tilde{g}^a + g^b \rightarrow \tilde{g}^c + \saxion$ (Crossing of D)
\item [Process F] $\tilde{q}_i + \bar{\tilde{q}}_j \rightarrow g^a + \saxion$
\begin{center}
\includegraphics[width=0.13\textwidth]{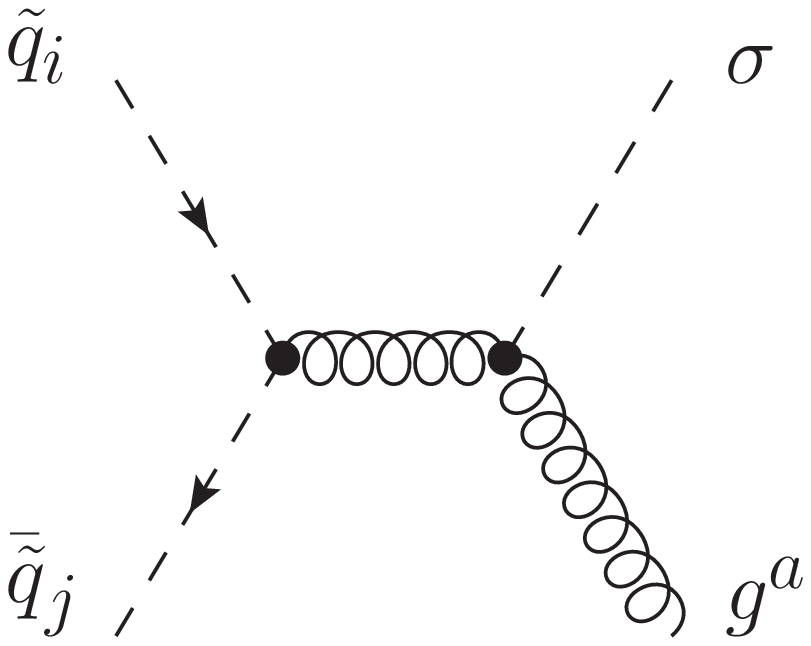}
\end{center}
\item [Process G] $\tilde{q}_i + g^a \rightarrow \tilde{q}_j + \saxion$ (Crossing of F)
\item [Process H] $\tilde{q}_i + \bar{q}_j \rightarrow \tilde{g}^a + \saxion$
\begin{center}
\includegraphics[width=0.13\textwidth]{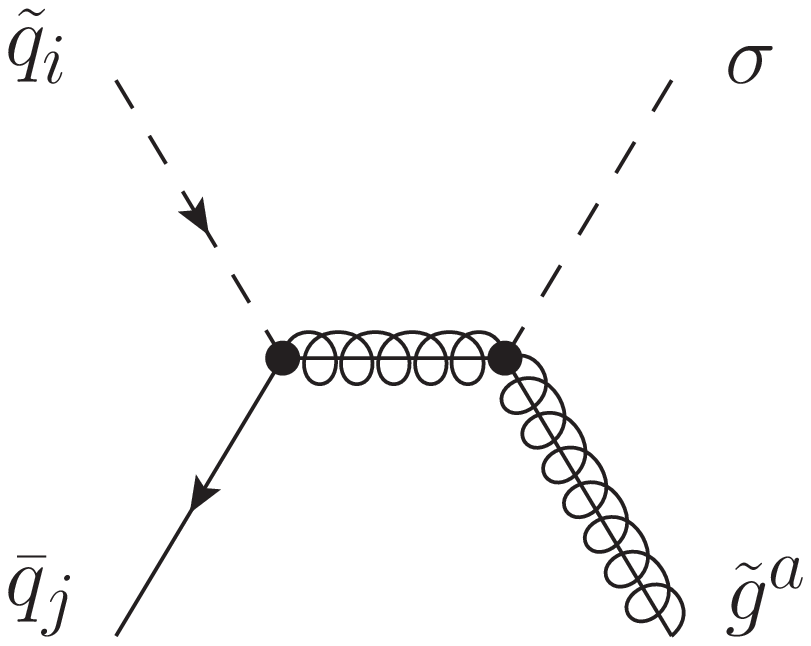}
\end{center}
\item [Process I] $\tilde{g}^a + q_i \rightarrow \tilde{q}_j + \saxion$ (Crossing of H)
\item [Process J] $\tilde{g}^a + \tilde{q}_i \rightarrow q_j + \saxion$ (Crossing of H)
\end{description}
\caption{{The $2\to2$ processes for saxion production in a SUSY QCD plasma. 
Additional processes are included in terms
of multiplicities in our calculation of the thermal production rate:
Process C with antiquarks $\bar{q}_{i,j}$ replacing $q_{i,j}$, 
process G with antisquarks $\bar{\tilde{q}}_{i,j}$ replacing $\tilde{q}_{i,j}$, 
process H with antisquarks/quarks $\bar{\tilde{q}}_{i}$/$q_{j}$ 
replacing $\tilde{q}_{i}$/$\bar{q}_{j}$,
and processes I and J with $\bar{q}_{i}$ and $\bar{\tilde{q}}_j$ 
replacing $q_{i}$ and $\tilde{q}_j$, respectively.}}
\label{Fig:harddiag}
\end{figure}

The calculation of the thermal production of saxions with $E\gtrsim T$ follows 
closely~\cite{Graf:2010tv}, where thermal axion production in a SM QGP is considered. 
%
\begin{table}[b]
\centering
\caption{Squared matrix elements for saxion ($\saxion$) production in 2-body processes involving MSSM quarks and squarks of a single chirality ($q_i$, $\tilde{q}_i$), gluons ($g^a$), and gluinos ($\tilde{g}^a$) in the high-temperature limit, $T \gg m_i$, with the SU(3)$_c$ color matrices $f^{abc}$ and $T^a_{ji}$. Sums over initial and final state spins have been performed.}
\label{Tab:Misquared}
\begin{ruledtabular}
\begin{tabular}{@{\extracolsep{\fill}}ccc}
Label $i$ 
& Process $i$ 
& $|M_i|^2 / \left(\frac{g_s^6}{128\pi^4 \fax^2}\right)$ \\ 
\noalign{\smallskip}
\hline
\noalign{\smallskip}
A 
& $g^a + g^b \rightarrow g^c + \saxion$ 
& $-4\frac{(s^2+st+t^2)^2}{st(s+t)}|f^{abc}|^2$ \\
B 
& $q_i + \bar{q}_j \rightarrow g^a + \saxion$ 
& $\left(\frac{2t^2}{s} + 2t+s\right)|T_{ji}^a|^2$ \\
C 
& $q_i + g^a \rightarrow q_j + \saxion$
& $\left(-\frac{2s^2}{t}-2s-t\right)|T_{ji}^a|^2$ \\
D
&$\tilde{g}^a + \tilde{g}^b \rightarrow g^c + \saxion$
&$2\left(\frac{2t^2}{s} + 2t+s\right)|f^{abc}|^2$ \\
E
&$\tilde{g}^a + g^b \rightarrow \tilde{g}^c + \saxion$
&$2\left(-\frac{2s^2}{t}-2s-t\right)|f^{abc}|^2$\\
F
&$\tilde{q}_i + \bar{\tilde{q}}_j \rightarrow g^a + \saxion$
&$-2\left(\frac{t^2}{s}+t\right)|T_{ji}^a|^2$\\
G
&$\tilde{q}_i + g^a \rightarrow \tilde{q}_j + \saxion$
&$-2\left(\frac{s^2}{t}+s\right)|T_{ji}^a|^2$\\
H
&$\tilde{q}_i + \bar{q}_j \rightarrow \tilde{g}^a + \saxion$
&$(s+t)|T_{ji}^a|^2$\\
I
&$\tilde{g}^a + q_i \rightarrow \tilde{q}_j + \saxion$
&$s|T_{ji}^a|^2$\\
J
&$\tilde{g}^a + \tilde{q}_i \rightarrow q_j + \saxion$
&$-t|T_{ji}^a|^2$\\
\end{tabular}
\end{ruledtabular}
\end{table}
%
From~\eqref{Eq:eff_lag} we get the relevant $2\to2$ processes 
shown in Fig.~\ref{Fig:harddiag}.%
\footnote{Note that $2\to2$ processes,  
such as $g^a + g^b\to\saxion+\axion$,
which involve the saxion-(s)axion interaction~\eqref{Eq:axion_saxion}
are suppressed by an additional factor of $1/\fax^2$ in
the respective squared matrix element and thus negligible.
Other processes
that involve saxions and/or axions in the initial state are
suppressed since their contribution to the rate is proportional to
the saxion/axion phase space density $f_{\saxion/\axion}$.
The latter is much smaller than the equilibrium densities
of the colored particles in the hot plasma
when $T$ is well below the saxion/axion decoupling temperature $\TD$.}
Additional processes exist 
but can be accounted for by multiplying the squared matrix elements
of the shown processes with appropriate multiplicity factors.
The squared matrix elements of the shown processes are listed in
Table~\ref{Tab:Misquared}, where $s=(P_1+P_2)^2$ and 
$t=(P_1-P_3)^2$ with $P_1$, $P_2$, $P_3$, and $P$ 
referring to the particles in the given order. 
Working in the limit, $T \gg m_i$, 
the masses $m_i$ of all MSSM particles involved have been neglected.  
Sums over initial and final spins have been performed. 
For quarks and squarks, the contribution of a single chirality is given. 
The obtained squared matrix elements can be calculated conveniently, 
e.g., with the help of {\tt FeynArts}~\cite{Hahn:2000kx}
and {\tt FormCalc}~\cite{Hahn:1998yk}. 

The results for processes A, C, E, and G 
given in Table~\ref{Tab:Misquared}
point to potential infrared (IR) divergences. 
Here screening effects of the plasma become relevant. 
In Refs.~\cite{Braaten:1989mz,Braaten:1991dd} 
a systematic method is introduced
to account for such screening effects in a gauge-invariant way. 
Following Ref.~\cite{Braaten:1991dd}, we introduce a momentum scale
$\kcut$ such that $g_sT\ll\kcut\ll T$ in the weak coupling limit
$g_s\ll 1$. This separates soft gluons with momentum transfer of order
$g_sT$ from hard gluons with momentum transfer of order $T$. By
summing the respective soft and hard contributions, the finite rate
for thermal production of saxions with $E\gtrsim T$ is obtained in
leading order in $g_s$, 
\begin{align}
E\frac{dW\sax}{d^3p} 
=
E\frac{dW\sax}{d^3p}\biggr\vert_{\text{soft}} 
+
E\frac{dW\sax}{d^3p}\biggr\vert_{\text{hard}},
\label{Eq:AxionTPRate}
\end{align}
which is independent of $\kcut$. 

\begin{figure}[t]
\centering
\includegraphics[width=0.30\textwidth]{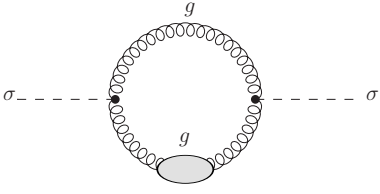}
\caption{{The saxion self energy used to compute 
the leading contribution to the thermal production rate
of hard saxions. 
The blob indicates the HTL-resummed gluon propagator.}}
\label{Fig:softdiag}
\end{figure}
%
%
In the region where $k<\kcut$, we use the optical theorem 
to obtain the soft contribution 
from the imaginary part of the saxion self energy 
shown in Fig.~\ref{Fig:softdiag}. 
Since only one gluon can carry a soft momentum, 
we need to use the HTL-resummed propagator only once. 
Using $\kcut$ as the ultraviolet cutoff, we get
\begin{align}
&E\frac{dW\sax}{d^3p}\biggr\vert_{\text{soft}} 
= 
-\frac{\fb(E)}{(2\pi)^3}
\operatorname{Im}\Pi_{\saxion}(E+i\epsilon,\vec{p})\vert_{k<\kcut}
\label{Eq:OpticalTheorem}
\\
&=
E\fb(E)\frac{3m_g^2g_s^4(N_c^2-1)T}{8192\pi^8 \fax^2}\!\!
\left[ \ln\!\!\left( \frac{\kcut^2}{m_g^2} \right) -1.379 \right],
\label{Eq:SoftPart}
\end{align}
where the squared SUSY thermal gluon mass is given by
$m_g^2=g_s^2T^2(N_c+n_f)/6$ 
for $N_c=3$ colors and $n_f=6$ light quark flavors 
and the equilibrium phase space density for bosons (fermions) by
$f_{\mathrm{B(F)}}(E)=[\exp(E/T)\mp 1]^{-1}$. 
More details on the way in which this calculation is performed can be found in 
Refs.~\cite{Braaten:1991dd,Bolz:2000fu,Pradler:thesis,Graf:thesis}.

In the region where $k>\kcut$ we can use zero temperature Feynman rules since $\kcut$ provides an IR cutoff. From the matrix elements given in Table~\ref{Tab:Misquared}, weighted with appropriate multiplicities, statistical factors, and phase space distributions, we get the (angle-averaged) hard contribution
\begin{align}
& 
E\frac{dW\sax}{d^3p}\biggr\vert_{\text{hard}}
=
\frac{1}{2(2\pi)^3}
\int\frac{d\Omega_p}{4\pi} 
\int\left[ \prod_{j=1}^3\frac{d^3p_j}{(2\pi)^3 2E_j} \right]  
\nonumber\\
&\times(2\pi)^4 \delta^4(P_1+P_2-P_3-P)\Theta(k-\kcut)
\nonumber\\
&\times \sum f_1(E_1)f_2(E_2)[1\pm f_3(E_3)] | M_{1+2\rightarrow 3+\sigma} | ^2
\label{Eq:HardPartStart}
\\
&
=E\frac{g_s^6(N_c^2-1)(N_c+n_f)}{512\pi^7 \fax^2} 
\left\{
\frac{\fb(E)T^3}{48\pi}\ln(2)
\right. 
\nonumber
\\
&
+\frac{\fb(E)T^3}{32\pi}
\left[ \ln \left(\frac{T^2}{\kcut^2}\right) + \frac{17}{3} - 2\gamma + \frac{2\zeta'(2)}{\zeta(2)} \right] 
\label{Eq:HardPartResult}
\\
&
\left. 
+(I_{\text{BBB}}^{(1)}+I_{\text{FBF}}^{(1)}-I_{\text{BBB}}^{(3)}+I_{\text{FFB}}^{(3)})
-2\frac{n_f(I_{\text{FBF}}^{(2)}+I_{\text{FFB}}^{(2)})}{N_c+n_f} 
\right\}
\nonumber
\end{align}
with Euler's constant $\gamma$, Riemann's zeta
function $\zeta(z)$,
\begin{align}
&
I_{\text{BBB(FBF)}}^{(1)}
= 
\frac{1}{32\pi^3 E^2}
\!\int_0^\infty \! dE_3 
\!\int_0^{E+E_3} \! dE_1
\ln\left(\frac{|E_1-E_3|}{E_3}\right) 
\nonumber\\
&  
\times
\left\{- \Theta(E_1-E_3)
\frac{d}{dE_1}\!\!
\left[f_{\text{BBB(FBF)}}E_2^2(E_1^2+E_3^2)\right] 
\right. 
\nonumber\\
& 
\,\,\,\,
+ \Theta(E_3-E_1)
\frac{d}{dE_1}
[f_{\text{BBB(FBF)}} E^2 (E_1^2+E_3^2)] 
\nonumber\\
& 
\left. 
\,\,\,\,
+ \Theta(E-E_1)
\frac{d}{dE_1}\!\!
\left[f_{\text{BBB(FBF)}}\!\!\left(E_1^2E_2^2-E_3^2 E^2\right)\right] 
\right\},
\label{Eq:I1}
\\
&
I_{\text{FBF(FFB)}}^{(2)}
= 
\frac{1}{96\pi^3 E^2}
\int_0^\infty dE_3 
\int_0^{E+E_3} dE_2 \, f_{\text{FBF(FFB)}}
\nonumber \\
&
\times
\left\{
\Theta(E-E_3)\left[E_3^2\left(E_3-3E_1\right)
	-\Theta(E_2-E)(E_2-E)^3\right]
\right.
\nonumber \\
& 
\,\,\,\,
+ \Theta(E_3-E)\left[\left(E-3E_2\right)E^2 
	+ \Theta(E-E_2)(E_2-E)^3\right] 
\nonumber \\
&
\,\,\,\,
+\left[\Theta(E_3-E_2)\Theta(E-E_3)-\Theta(E_3-E)\Theta(E_2-E_3)\right]
\nonumber \\
&
\quad\times
\left.\left[(E_2-E_3)^2(E_2+2E_3)-3(E_2^2-E_3^2)E\right]
\right\},
\label{Eq:I2} 
\\
&
I_{\text{BBB(FFB)}}^{(3)} 
= 
\frac{1}{32\pi^3 E^2} 
\int_0^\infty dE_3 
\int_0^{E+E_3} dE_2 \, f_{\text{BBB(FFB)}} 
\nonumber \\
&
\times 
\left\{ 
\Theta(E-E_3) \frac{E_1^2E_3^2}{E_3+E} 
+ 
\Theta(E_3-E) \frac{E^2 E_2^2}{E_3+E}
\right. 
\nonumber\\
& 
\quad\,
+ \! [\Theta(E_3-E)\Theta(E_2-E_3)-\Theta(E-E_3)\Theta(E_3-E_2)]
\nonumber\\
& 
\quad\quad
\times (E_2-E_3)\,[E_2(E_3-E)-E_3(E_3+E)] 
\biggr\},
\label{Eq:I3}
\\
& 
f_{\text{BBB,FBF,FFB}}
=
f_1(E_1)f_2(E_2)[1\pm f_3(E_3)] 
.
\label{Eq:fXXX}
\end{align}
The sum in~\eqref{Eq:HardPartStart} is over all saxion production
processes $1+2\rightarrow 3+\saxion$ viable
with~\eqref{Eq:eff_lag}. The colored particles 1--3 were in
thermal equilibrium at the relevant times. Performing the calculation
in the rest frame of the plasma, $f_{i}$ are thus described by
$f_{\mathrm{F/B}}$ depending on the respective spins. Shorthand
notation~\eqref{Eq:fXXX} indicates the corresponding combinations,
where $+$ ($-$) accounts for Bose enhancement (Pauli blocking) when
particle~$3$ is a boson (fermion). 
With any initial saxion population diluted away by inflation 
and for $T$ well below
the saxion decoupling temperature $\TD$ 
(which will be determined in Sect.\ref{Eq:SaxionAxionYield}), 
we can neglect saxion disappearance reactions 
and Bose enhancement by saxions, 
since the saxion phase space density 
$f\sax\ll f_{\mathrm{F/B}}$ and $1+f\sax \approx1$.

\section{Thermal axion production}
\label{Sec:ThermalAxionProduction}

The calculation of thermal axion production 
in the primordial SUSY QCD plasma
proceeds analogously to the saxion calculation 
presented in the previous section.
After substituting the saxion $\sigma$ by the axion $a$, 
the Feynman diagrams can be read directly 
from Figs.~\ref{Fig:harddiag} and~\ref{Fig:softdiag} 
with one modification: 
there is no gluino-gluino-gluon-axion vertex and thus no quartic interaction 
such as the one that contributes to processes~D and~E in the saxion case.

Although the Feynman rules for the axion interactions 
derived from~\eqref{Eq:eff_lag} differ 
from the ones describing saxion interactions, 
we obtain squared matrix elements for the axion production processes
in the high-temperature limit, $T \gg m_i$,
that agree with the ones for the corresponding saxion production processes
given in Table~\ref{Tab:Misquared}.
Moreover, we find that both the soft and the hard contributions to the
thermal production rate of hard axions agree with~\eqref{Eq:SoftPart} 
and~\eqref{Eq:HardPartResult}, respectively.
Our result for the thermal axion production rate $E\,dW\ax/d^3p$ 
thus agrees with the one for the thermal saxion production rate obtained above.
This implies an agreement of the associated thermally produced yields 
of axions and saxions prior to decay, which will be calculated in the next section.

Before proceeding let us stress that we can neglect production processes
like $\pi\pi\to\pi\axion$ in the primordial hot hadronic gas~\cite{Chang:1993gm,Hannestad:2005df} because of the $\fax$ limit~\eqref{Eq:fa_Limit}.
Moreover, Primakoff processes such as $\electron\gamma\to\electron\axion$ 
are not taken into account since they are usually far less efficient in
the early Universe~\cite{Turner:1986tb}.

\section{Thermal saxion/axion yield}
\label{Eq:SaxionAxionYield}

Let us now calculate the thermally produced
(TP) saxion yield $\YsaxTP=\nsax/s$, where $\nsax$ is the corresponding
saxion number density and $s$ the entropy density.
With the results obtained in the two previous sections, 
we know beforehand that this yield prior to decay agrees
with the thermally produced axion yield $\YaxTP=\nax/s$.
While the calculation and results are indeed valid for both saxion and axion,
we focus on the saxion case. 

For $T$ sufficiently below the saxion decoupling temperature $\TD$, 
the evolution of the thermally produced $\nsax$ with cosmic time $t$ 
is governed by the Boltzmann equation
\be 
\frac{d\nsax}{dt} + 3H\nsax 
= 
\int d^3p \,\frac{dW\sax}{d^3p}
=
W_{\saxion} .
\label{Eq:BoltzmannEquation}
\ee
Here $H$ is the Hubble expansion rate, and the collision term is the
integrated thermal production rate:
\be
W_{\saxion} 
= 
\frac{9\zeta(3) g_s^6 T^6}{256\pi^7\fax^2} 
\left[ \ln\left(\frac{T^2}{m_g^2} \right) + 0.4305 \right] .
\label{collisionSM}
\ee
Assuming conservation of entropy per comoving volume element,
$\eqref{Eq:BoltzmannEquation}$ can be written as
$d\YsaxTP/dt=W_{\saxion}/s$.
Since thermal saxion production is efficient only in the 
hot radiation dominated epoch with temperatures well above 
the one of radiation-matter equality, $T_\mathrm{mat=rad}$,
we can change variables from cosmic time $t$ to temperature $T$ accordingly.
With an initial temperature $\TR$ at which $\YsaxTP(\TR)\simeq 0$, 
the relic saxion yield prior to decay is
\begin{align}
&\YsaxTP
\approx
\YsaxTP(T_\mathrm{low})
= \int_{T_\mathrm{low}}^{\TR} dT
\frac{W_{\saxion}(T)}{T s(T) H(T)}
\nonumber\\
&= 
1.33\times 10^{-3} g_s^6 
\ln \! \left(\frac{1.01}{g_s}\right) \!
\left( \frac{10^{11}\,\GeV}{\fax} \right)^{\!\!2} \!
\left( \frac{\TR}{10^{8}\,\GeV} \right)\! 
\label{Eq:SAxionYieldTP}
\end{align}
with a fiducial temperature $T_\mathrm{low}$ well below $\TR$ 
and well above $T_\saxion$, which we use to denote
the temperature of the primordial plasma at $t=\tausaxion$:
$T_\saxion\ll T_\mathrm{low}\ll \TR$.
In the case of the axion, 
$T_\mathrm{low}=T_\mathrm{mat=rad}$ can be used 
since its lifetime exceeds the time of radiation-matter equality significantly.
Note that the resulting saxion/axion yield is insensitive 
to the exact choice of $T_\mathrm{low}$ for $T_\mathrm{low}<0.01\,\TR$
since additional contributions from thermal production 
at $T<0.01\,\TR$ are found to be negligible.

\begin{figure}[t]
\centering
\includegraphics[width=0.45\textwidth,clip=true]{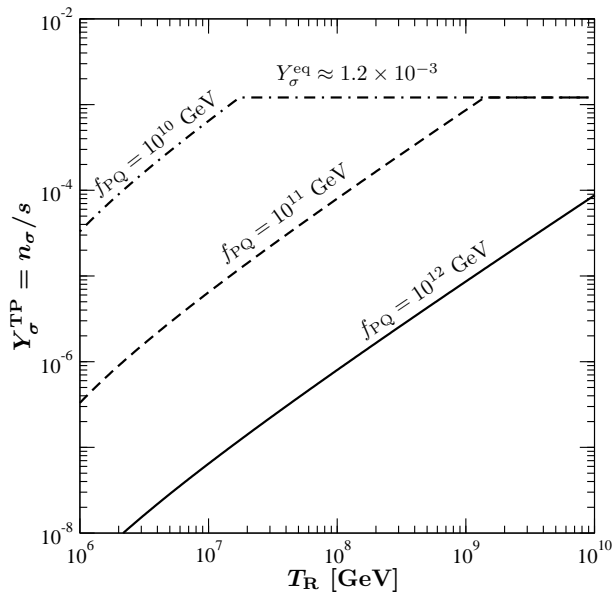}
\caption{{The relic saxion yield prior to decay originating from thermal
  processes in the primordial plasma for cosmological scenarios
  characterized by different $\TR$ values covering the range from
  $10^6$ to $10^{10}\,\GeV$. The dash-dotted, dashed, and
  solid lines are obtained for $\fax=10^{10}$, $10^{11}$, and
  $10^{12}\,\GeV$. The relic axion yield $\YaxeqTP$ from thermal processes
  agrees with $\YsaxeqTP\!\!$ and can thus be read from this figure as well.}}
\label{Fig:yield}
\end{figure}

Figure~\ref{Fig:yield} shows the saxion yield~\eqref{Eq:SAxionYieldTP}
for $\fax = 10^{10}$, $10^{11}$, and $10^{12}\,\GeV$ 
as the diagonal dash-dotted, dashed, and solid lines, respectively.
%
%
%
Here we compute~\eqref{Eq:SAxionYieldTP} 
with $g_s\equiv g_s(\TR)=\sqrt{4\pi\alpha_s(\TR)}$ evaluated 
according to its 1-loop renormalization group running within the MSSM
from $\alpha_s(m_\mathrm{Z})=0.1176$ at the Z-boson mass
$m_\mathrm{Z}=91.1876~\GeV$.
The applied methods~\cite{Braaten:1989mz,Braaten:1991dd} 
require $g_s\ll1$, 
so that~\eqref{Eq:SAxionYieldTP} is most reliable for $\TR\gg10^6\,\GeV$. 
For lower $\TR$ values such that $g_s(\TR)\gtrsim 1$,
one encounters an artificial suppression of $\YsaxTP$
and even unphysical negative values,
which can be seen directly from the logarithmic factor 
in~\eqref{Eq:SAxionYieldTP}.
This is a well-known limitation of this technique (cf.~\cite{Brandenburg:2004du,Graf:2010tv})
that calls for generalizations 
of the gauge-invariant methods 
introduced in Refs.~\cite{Braaten:1989mz,Braaten:1991dd}
modified to extend the applicability beyond the weak coupling limit.

Note that~\eqref{Eq:SAxionYieldTP} is only valid if $\TR\ll\TD$, 
because otherwise saxion annihilation processes 
neglected in~\eqref{Eq:HardPartStart} are important. 
For $\TR\gtrsim\TD$ saxions were in thermal equilibrium 
in the early Universe before decoupling as thermal relics. 
In fact, for $\msaxion\ll\TD$,
saxions decouple as a relativistic species.
The yield is then given by
\be
Y\sax^\eq = \frac{n^\eq_{\saxion}}{s}\approx1.2\times10^{-3}
\label{Eq:eqyield}
\ee
as indicated by the horizontal lines in Fig.~\ref{Fig:yield}. 
The yield from thermal production cannot exceed the equilibrium yield, 
so that~\eqref{Eq:eqyield} represents an upper limit.
In scenarios with values of $\TR$ 
for which~\eqref{Eq:SAxionYieldTP} is close to 
or larger than~\eqref{Eq:eqyield}, 
saxion disappearance reactions have to be taken into account.%
\footnote{Here also the additional processes 
$\saxion\leftrightarrow\axion\axion$ can become efficient 
that involve the saxion-axion interaction~\eqref{Eq:axion_saxion}
governed by $x$.}
The resulting yield would show almost the same $\TR$ dependence 
as the one in Fig.~\ref{Fig:yield}, 
but with a smooth transition in the $\TR$ range 
in which $\YsaxTP \simeq Y\sax^\eq$. 

For axions, $\maxion\ll\TD$ is always satisfied so that 
they will be hot thermal relics 
with $\Yaxeq=Y\sax^\eq$ as given in~\eqref{Eq:eqyield} if $\TR\gtrsim\TD$.
In fact, the relic axion yield $\YaxeqTP$ from thermal processes
agrees fully with $\YsaxeqTP\!\!$ described above 
and can be read from Fig.~\ref{Fig:yield} as well.

\section{Decoupling Temperature}
\label{Sec:TD}

Considering Fig.~\ref{Fig:yield}, one finds that 
the kinks indicate critical $\TR$ values. 
For a given $\fax$, the associated critical $\TR$ value separates scenarios 
with thermal relic saxions from those in which saxions have never been in thermal equilibrium.
We thus use the positions of the kinks
as $\fax$-dependent estimates of the saxion decoupling temperature. 
Our numerical results are well described by
\be
\TD
\approx 
1.4\times 10^9\text{ GeV }\left( \frac{\fax}{10^{11}\text{ GeV}} \right) ^{2}.
\label{Eq:TD}
\ee
This is similar to the estimate of the axino decoupling temperature 
in Ref.~\cite{Rajagopal:1990yx}.
Such an agreement was expected and used to provide estimates of 
the thermally produced saxion yield $\YsaxTP$ 
in Refs.~\cite{Asaka:1998ns,Asaka:1998xa,Asaka:2000ew,Kawasaki:2007mk}.
Another recent study applies the thermally produced axino yield 
$Y_{\tilde{a}}^\TP$~\cite{Brandenburg:2004du}
to estimate
$\YsaxTP
\simeq
(2/3)\,Y_{\tilde{a}}^\TP$~\cite{Moroi:2012vu}.%
\footnote{Note that $\fax$ in~\cite{Asaka:1998ns,Asaka:1998xa}
and $F_a$ in~\cite{Kawasaki:2007mk,Moroi:2012vu}
correspond to our $\vax=\fax/\sqrt{2}$ 
and thereby differ by $1/\sqrt{2}$ from our $\fax$. 
With these differences in the definitions of the PQ scale, 
we find that the $\YsaxTP$ estimates 
in Refs.~\cite{Asaka:1998ns,Asaka:1998xa,Kawasaki:2007mk,Moroi:2012vu} 
exceed the result~\eqref{Eq:SAxionYieldTP} of our calculation 
by about a factor of two for fixed $\fax$ and $\TR$.}
With our results illustrated in Fig.~\ref{Fig:yield} above,
one can now see explicitly the similarity between $\YsaxTP$ 
and the corresponding axino yield $Y_{\tilde{a}}^\TP$ 
illustrated in Fig.~4 of Ref.~\cite{Brandenburg:2004du}.

In light of Sect.~\ref{Sec:ThermalAxionProduction}, 
it is clear that~\eqref{Eq:TD} describes
the axion decoupling temperature in the considered SUSY settings
as well.
When comparing~\eqref{Eq:TD} 
with the axion decoupling temperature in non-SUSY scenarios, 
given in Eq.~(15) of Ref.~\cite{Graf:2010tv},
we find only small differences.
In fact, for a fixed $g_s\ll 1$, the additional diagrams in the SUSY case 
(which lead to a different thermal gluon mass $m_g$ also)
increase the collision term for thermal axion production $W_a$ 
only by at most 30\%  with respect to Eq.~(12) of Ref.~\cite{Graf:2010tv} 
obtained for the non-SUSY case.
Note also that both $\YaxTP$ and $Y\ax^\eq$ 
are normalized to an entropy density $s(T)$
which is more than two times larger in the SUSY case than in the non-SUSY case
due to the additional sparticles which can all be considered to be relativistic
at very high temperatures such as the axion decoupling temperature.

\section{Additional radiation from saxion decays}
\label{Sec:AddRad}

As already mentioned in the Introduction, axions from late saxion decays
can provide additional radiation already prior to BBN and later on as well.
The amount of additional radiation is usually expressed 
in terms of a non-standard contribution $\Delta\neff$
to the effective number of light thermally excited neutrino species \neff. 
It is defined in relation 
to the total relativistic energy density
\be
\rho_\text{rad}(T)
=
\left[ 1+\frac{7}{8}\neff\left(\frac{T_\nu}{T}\right)^4\right] 
\rho_\gamma(T)
\label{Eq:neff_def}
\ee
with the photon energy density $\rho_\gamma$ and
the temperatures of neutrinos $T_\nu$
and of photons $T$. 
At $T\gtrsim 1~\MeV$ (before neutrino decoupling and $e^+e^-$ annihilation),
$\neff=3+\Delta\neff$ and $T_\nu=T$.
These relations change 
to $T_\nu=(11/4)^{-1/3}T$ after neutrino decoupling
and to $\neff=3.046+\Delta\neff$~\cite{Mangano:2005cc} 
because of residual neutrino heating by $e^+e^-$ annihilation.

At a photon temperature $T<T_\saxion$,
the energy density of relativistic non-thermally produced (NTP) axions 
from saxion decays $\rho_{\axion}^{\NTP}(T)$ yields
\be
\Delta\neff(T) 
= 
\frac{120}{7\pi^2 T_\nu^4}\, \rho_{\axion}^{\NTP}(T).
\label{Eq:DeltaNeffAxionNTP}
\ee
Working in the sudden decay approximation, 
all thermally produced saxions are considered
to decay instantaneously 
at $t=\tausaxion$ (where $T=T_\saxion$). 
If the saxions are non-relativistic when decaying dominantly into two axions,
the initial axion momentum is $p_\axion(T_\saxion)=\msaxion/2$ and
\begin{eqnarray}
\rho_{\axion}^{\NTP}(T)
&=&
\frac{m_\saxion}{2}
\left[\frac{g_{*S}(T)}{g_{*S}(T_\saxion)}\right]^{1/3}
\frac{T}{T_\saxion}\,
Y_\axion^\NTP s(T)
\\
&=& 
\left[\frac{g_{*S}(T)}{g_{*S}(T_\saxion)}\right]^{4/3}
\left(\frac{T}{T_\saxion}\right)^4\,
\rho_\saxion^{\eq/\TP}(T_\saxion) 
\label{Eq:rhoax2nd}
\end{eqnarray}
with 
$\rho_\saxion^{\eq/\TP}\!(T_\saxion)
=
m_\saxion Y_\saxion^{\eq/\TP}\!s(T_\saxion)$ 
and
$Y_\axion^\NTP\!=2 Y_\saxion^{\eq/\TP}$.
Here $g_{*S}$ denotes the number of effectively massless degrees of freedom
such that $s=2\pi^2 g_{*S} T^3/45$.
For $\tausaxion\simeq1/\Gamma_{\saxion\to aa}$ given by~\eqref{Eq:sax_lifetime}
and with the time-temperature relation 
in the radiation-dominated epoch, 
we obtain
\be
T_\saxion
\simeq
10.6~\MeV 
\left(\frac{\msaxion}{1~\GeV}\right)^{\!3/2}\!
\left(\frac{10^{10}\,\GeV}{\fax/x}\right)\!
\left[\frac{10.75}{g_{*}(T_\saxion)}\right]^{1/4}
\label{Eq:Tsaxion}
\ee
and
\begin{eqnarray}
\Delta\neff(T) 
&\simeq& 
\frac{0.95}{x}
\left(\frac{1~\GeV}{\msaxion}\right)^{\!\!1/2} \!
\left(\frac{\fax}{10^{10}\,\GeV}\right) \!
\left(\frac{Y_\saxion^{\eq/\TP}}{10^{-3}}\right)
\nonumber\\
&&
\times
\left(\frac{T}{T_\nu}\right)^4
\left[\frac{g_{*S}(T)}{10.75}\right]^{4/3} \!
\frac{g_{*}(\tausaxion)^{1/4}}{g_{*S}(\tausaxion)^{1/3}},
\label{Eq:DeltaNeffYield}
\end{eqnarray}
where $g_{*}$ is the effective number of relativistic degrees of freedom
governing the energy density. 
Note that our focus on scenarios in which
saxions decay predominantly into axions
implies that~\eqref{Eq:Tsaxion}, \eqref{Eq:DeltaNeffYield}, 
and related expressions given below are valid only
down to $\msaxion$-dependent minimum values of $x$
as discussed at the end of Sect.~\ref{Sec:Setting}.

Focussing on saxions from thermal processes, the maximum $\Delta\neff$ 
emerges for scenarios with $\TR$ above the decoupling temperature~\eqref{Eq:TD}
so that the thermal relic yield~\eqref{Eq:eqyield} applies:
\begin{eqnarray}
\Delta\neff(T)
&\simeq& 
\frac{1.14}{x}
\left(\frac{1~\GeV}{\msaxion}\right)^{\!1/2} \!
\left(\frac{\fax}{10^{10}\,\GeV}\right)
\nonumber\\
&&
\times
\left(\frac{T}{T_\nu}\right)^4
\left[\frac{g_{*S}(T)}{10.75}\right]^{4/3} \!
\frac{g_{*}(\tausaxion)^{1/4}}{g_{*S}(\tausaxion)^{1/3}}.
\label{Eq:DeltaNeff1}
\end{eqnarray}
For $\TR<\TD$ on the other hand, 
the yield~\eqref{Eq:SAxionYieldTP} leads to:
\begin{eqnarray}
\Delta\neff(T)
&\simeq& 
\frac{12.6 g_s^6 \ln\left(\frac{1.01}{g_s}\right)}{x}
\left(\frac{1~\GeV}{\msaxion}\right)^{\!1/2} \!
\left(\frac{10^{10}\,\GeV}{\fax}\right) \!
\nonumber\\
&&
\hskip -2cm
\times
\left( \frac{\TR}{10^{7}\,\GeV} \right)
\left(\frac{T}{T_\nu}\right)^4
\left[\frac{g_{*S}(T)}{10.75}\right]^{4/3} \!
\frac{g_{*}(\tausaxion)^{1/4}}{g_{*S}(\tausaxion)^{1/3}}.
\label{Eq:DeltaNeff2}
\end{eqnarray}

Thermal relic saxions are non-relativistic when decaying 
if their average momentum at $T_\saxion$ satisfies
\be
\langle p(T_\saxion) \rangle
=
\langle p(\TD) \rangle
\left[\frac{g_{*S}(T_\saxion)}{g_{*S}(\TD)}\right]^{1/3}
\frac{T_\saxion}{\TD}
\ll
\msaxion .
\label{Eq:SaxionEqNR1}
\ee
Since those saxions decouple as a relativistic species 
(provided $\msaxion\ll\TD$)
at a very high temperature~\eqref{Eq:TD} with a thermal spectrum,
$\langle p(\TD) \rangle=2.701\,\TD$ and $g_{*S}(\TD)\simeq 232.5$.%
\footnote{This $g_{*S}$ value accounts for the MSSM and
the axion multiplet fields, which can all be considered as relativistic at $\TD$
if not only $\msaxion$ but also the axino mass satisfies $\maxino\ll\TD$.}
Using~\eqref{Eq:Tsaxion},
we can express~\eqref{Eq:SaxionEqNR1}
in terms of the following 
$\msaxion$-dependent lower limit on the PQ scale
\be
\frac{\fax}{x}
\gg
8.4\times 10^7\,\GeV\,
\left(\frac{\msaxion}{1~\GeV}\right)^{\!1/2}
\frac{g_{*S}(T_\saxion)^{1/3}}{g_{*}(T_\saxion)^{1/4}}.
\label{Eq:SaxionEqNR3}
\ee
Almost the same limit applies to thermally produced saxions as well
since their production is efficient only at high temperatures
not far below $\TR$ and leads basically to a thermal spectrum,
i.e., \eqref{Eq:SaxionEqNR1} applies after substituting $\TD$ by $\TR$
and $g_{*S}(\TD)$ by $g_{*S}(\TR)\simeq 228.75$.

Note that the saxions decay while being decoupled from the primordial plasma 
if $T_\saxion\ll\TD$ or equivalently
\be
\frac{\fax}{x^{1/3}}
\gg
7.1\times 10^6\,\GeV
\left(\frac{\msaxion}{1~\GeV}\right)^{\!1/2}\!
\left[\frac{232.5}{g_{*}(T_\saxion)}\right]^{1/12}.
\label{Eq:faxTsaxionTD}
\ee
If this condition is satisfied the axions emitted in those decays
will not be thermalized but free-streaming.
Thus, the temperature $\TNR$ at which the 
non-thermally produced axions become non-relativistic reads
\begin{eqnarray}
\TNR
&=&
0.15\,x~\eV
\left(\frac{\msaxion}{10~\TeV}\right)^{1/2}
\left(\frac{10^{9}\,\GeV}{\fax}\right)^{2}
\nonumber\\
&&
\times
\left[\frac{3.91}{g_{*S}(\TNR)}\right]^{1/3}
\frac{g_{*S}(T_\saxion)^{1/3}}{g_{*}(T_\saxion)^{1/4}},
\label{Eq:TNRaNTP}
\end{eqnarray}
when defined via $p_a(\TNR)=\maxion$.
This shows that the emitted axions are expected to be still relativistic
at the last scattering surface and even well thereafter 
for $\msaxion\lesssim 10~\TeV$ and $x=\Order(1)$.
Thereby they can contribute to $\Delta\neff$ even at late times 
where studies of the CMB and LSS
allow us to probe the amount of radiation.

\subsection{BBN}
\label{Sec:BBN}

For $T_\saxion\gtrsim 1~\MeV$, the axions from saxion decays 
contribute to the radiation density already at the onset of BBN 
and prior to $e^+e^-$ annihilation.
This leads to a speed-up of the Hubble expansion rate and thereby
to an output of $^4$He that is more efficient than
in standard BBN with $\Delta\neff=0$. 
In turn, the inferred primordial $^4$He abundance imposes 
upper limits on $\Delta\neff$, 
whereas the inferred primordial D abundance constrains the baryon density 
$\omega_\text{b}=\Omega_\text{b}h^2$,
with the normalized Hubble constant $h\simeq 0.7$. 
Notably, 
two recent studies of the primordial ${}^4$He mass fraction $Y_\text{p}$
even report values that point to an excess
over the standard BBN prediction:
Izotov and Thuan~\cite{Izotov:2010ca} find 
$Y^\text{IT}_\text{p} = 0.2565\pm0.001(\text{stat.})\pm0.005(\text{syst.})$
and Aver \textit{et al.}~\cite{Aver:2010wq} 
$Y^\text{Av}_\text{p} = 0.2561\pm0.0108$,
with all errors refering to 68\% intervals.
As mentioned in the Introduction, these results may be hints
towards extra radiation at the onset of BBN,
which can reside in the form of axions from decays
of saxions from thermal processes.

Based on recent studies of the primordial $^4$He 
and D abundances~\cite{Izotov:2010ca,Aver:2010wq,MNR:MNR13921} 
and the recent Particle Data Group (PDG) recommendation 
for the free neutron lifetime, 
$\tau_{\mathrm{n}}=880.1\pm1.1~\seconds$~\cite{Beringer}, 
we now derive $\Delta\neff$ limits 
from a BBN likelihood analysis, 
similar to the one in Ref.~\cite{Hamann:2011ge},
and explore the implications for the considered SUSY axion models.
Relying on the $Y_\text{p}$ results given above 
and on the primordial D abundance reported by 
Petini~\textit{et al.}~\cite{MNR:MNR13921},
$\log[\text{D/H}]_\text{p}=-4.56\pm0.04$,
we consider the two log-likelihood functions
\begin{eqnarray}
\ln L_{{}^4\text{He}} 
&=& 
\begin{cases}
-\frac{1}{2}\frac{(Y_\text{p} - 0.2565)^2}{0.0051^2} 
& \mathrm{for~[46]} 
\\
-\frac{1}{2}\frac{(Y_\text{p} - 0.2561)^2}{0.0108^2}
& \mathrm{for~[47]} 
\end{cases}
\label{Eq:LHe}\\
\ln L_{\text{D}} 
&=& 
-\frac{1}{2}\frac{(\log[\text{D/H}]_\text{p}+ 4.56)^2}{0.04^2},
\label{Eq:D}
\end{eqnarray}
where small uncertainties related to nuclear reaction rates and 
also the ones related to the free neutron lifetime $\tau_{\mathrm{n}}$ 
are not taken into account.
Theoretical values for the primordial ${}^4$He and D abundances 
are calculated with the BBN code {\tt PArthENoPE}~\cite{Pisanti:2007hk} 
using $\tau_\text{n}=880.1~\seconds$~\cite{Beringer} and 
$0\leq\Delta\neff\leq 4$ and $0.01\leq\omega_\text{b}\leq 0.03$ as flat priors.
Calculating the respective combined likelihood 
and after marginalizing over $\omega_\text{b}$,
we obtain for $\Delta\neff$
the maximum likelihood posteriors and the minimal 99.7\% credible intervals 
listed in the first two lines of Table~\ref{Tab:Neffconstrains}.%
\footnote{Note that the PDG-recommended value 
for the free neutron lifetime has changed recently
from $\tau_\text{n}=885.7\pm 0.8~\seconds$~\cite{Nakamura:2010zzi}
to $\tau_{\mathrm{n}}=880.1\pm1.1~\seconds$~\cite{Beringer}.
If we use $\tau_\text{n}=885.7~\seconds$ in {\tt PArthENoPE}, 
we can reproduce the posterior maxima and the minimal 95\% credible intervals 
given in the first two lines of Table~III in Ref.~\cite{Hamann:2010bk}.
In comparison, those posterior maxima are about 10\% below the values 
obtained with $\tau_\text{n}=880.1~\seconds$
given in our Table~\ref{Tab:Neffconstrains}.}
%
%
\begin{table}[t]
\centering
\caption{{Constraints on $\Delta\neff$ from BBN and precision cosmology. 
Based on the indicated data sets,
the first two lines give the posterior maximum (p.m.) 
and the minimal 99.7\% credible interval imposed by BBN
using the prior $\Delta\neff\geq 0$
and after marginalization over $\omega_\text{b}$.
The third line quotes the mean and the 95\% CL upper limit
on $\Delta\neff$ as obtained in the precision cosmology study
of Ref.~\cite{Hamann:2010pw} based on CMB data, 
the Sloan Digital Sky Survey (SDSS) data-release 7 halo power spectrum (HPS),
and data from the Hubble Space Telescope (HST).}}
\label{Tab:Neffconstrains}
\begin{ruledtabular}
\begin{tabular}{@{\extracolsep{\fill}}ccc}
Data
&  p.m./mean
&  upper limit\\
\noalign{\smallskip}
\hline
\noalign{\smallskip}
$Y_\text{p}^\text{IT}$~\cite{Izotov:2010ca} 
+ $[\text{D/H}]_\text{p}$~\cite{MNR:MNR13921}
& 0.76
& $<1.97~(3\sigma)$ \\
$Y_\text{p}^\text{Av}$~\cite{Aver:2010wq} 
+ $[\text{D/H}]_\text{p}$~\cite{MNR:MNR13921}
& 0.77
& $<3.53~(3\sigma)$ \\
CMB + HPS + HST~\cite{Hamann:2010pw}
& 1.73
& $<3.59~(2\sigma)$ \\
\end{tabular}
\end{ruledtabular}
\end{table}
%

\begin{figure}[t]
\centering
\includegraphics[width=0.45\textwidth,clip=true]{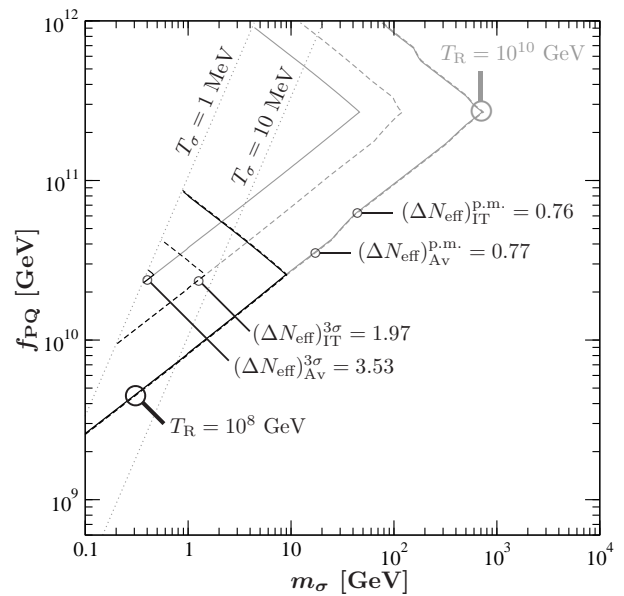}
\caption{{Contours of $\Delta\neff$ at $T\sim 1~\MeV$ 
provided by non-thermally produced axions from decays of
thermal saxions for $x=1$, $T_\saxion>1~\MeV$, and 
$\TR=10^8\,\GeV$ (black) and $10^{10}\,\GeV$ (gray).
The BBN results given in Table~\ref{Tab:Neffconstrains}
are illustrated by the solid (dashed) curves,
which indicate the posterior maximum 
$(\Delta\neff)_\text{Av(IT)}^\text{p.m.}=0.77$ $(0.76)$ 
and the upper $3\sigma$ limit 
$(\Delta\neff)_\text{Av(IT)}^{3\sigma}=3.53$ $(1.97)$
that disfavors the region to its left.
The dotted lines show $T_\saxion=1$ and $10~\MeV$.}}
\label{Fig:extraradBBN}
\end{figure}

Let us now apply these BBN constraints to the case of extra radiation 
from saxion decays into axions. 
We evaluate $\Delta\neff(T)$ from \eqref{Eq:DeltaNeff1} and~\eqref{Eq:DeltaNeff2} 
for $x=1$ and at $T\sim 1~\MeV$, 
i.e., at the onset of BBN and above the temperature at which neutrinos decouple.
Figure~\ref{Fig:extraradBBN} shows the resulting $\Delta\neff$ contours 
in the $m_\saxion$--$\fax$ parameter plane as black (gray) lines
for $\TR=10^8~(10^{10})\,\GeV$.
The solid (dashed) curves show -- as labeled -- the posterior maximum
$(\Delta\neff)_\text{Av(IT)}^\text{p.m.}=0.77$~$(0.76)$ 
and the upper limit
$(\Delta\neff)_\text{Av(IT)}^{3\sigma}=3.53\,(1.97)$,
which disfavors the considered region to its left by more than $3\sigma$.
The dotted lines indicate $T_\saxion=1$ and $10~\MeV$.
The parameter region with $T_\saxion<1~\MeV$ is not considered
since our BBN constraints on $\Delta\neff$ do not apply to later decays.%
\footnote{The calculations of {\tt PArthENoPE} 
start at $T=10~\MeV$
with the given $\Delta\neff$ values as input already at that temperature.
The $\Delta\neff$ limits derived above are thus strictly applicable 
for $T_\saxion\geq 10~\MeV$ only.
Modifications in the
{\tt PArthENoPE} code
that will allow us to describe more accurately 
the region with $T_\saxion<10~\MeV$
are postponed to future work.}
Moreover, as described in the Introduction, 
additional cosmological constraints
can occur for $T_\saxion<1~\MeV$,
which are beyond the scope of this work.
%

For $T_\saxion>1~\MeV$, 
one sees that the BBN constraints on $\Delta\neff$
can disfavor significant regions 
of the $m_\saxion$--$\fax$ parameter plane
in high $\TR$ scenarios.
These regions will become larger and move towards larger $\msaxion$ 
if $x$ is smaller than one but still 
sufficiently sizable such that the saxion decay into axions remains 
to be the dominant decay channel that governs $\tausaxion$.
Moreover, the shown posterior-maxima contours illustrate that 
non-thermally produced axions from decays of thermal saxions
can explain the existence of extra radiation, $\Delta\neff\sim 1$,
in agreement with the hints from BBN studies.
For $\TR>\TD$, the shape of the $\Delta\neff$ contours
is described by~\eqref{Eq:DeltaNeff1}
from decays of thermal relic saxions.
The kink of the $\Delta\neff$ contours indicates
the respective $\fax$ value at which $\TR=\TD$.
For larger $\fax$, $\TR<\TD$ and
\eqref{Eq:DeltaNeff2} applies which is reflected by
the $\TR$ dependence of $\Delta\neff$ provided
by axions from decays of thermally produced saxions.

As mentioned in the Introduction,
the energy density of the early Universe
can receive contributions not only from thermal saxions but also
from coherent oscillations of the saxion 
field~\cite{Chang:1996ih,Hashimoto:1998ua,Asaka:1998ns,Kawasaki:2007mk,Kawasaki:2011aa}.
In fact, axions from the decay of those non-thermal saxions 
constitute additional radiation as well~\cite{Kawasaki:2007mk,Kawasaki:2011aa}
and can thereby increase $\Delta\neff$ already at the onset of BBN.
However, in the parameter region considered in Fig.~\ref{Fig:extraradBBN},
their contribution to $\Delta\neff$ is basically negligible 
for an initial displacement of the saxion field from the vacuum
of $\sigma_{i}\sim\fax$.
This can be seen, for example, in Fig.~1 of Ref.~\cite{Kawasaki:2007mk}
where the energy density of coherent saxion oscillations
is compared with an estimate of the one of thermal saxions.

At this point, it is important to stress that significant additional restrictions
are possible that depend on the mass spectrum and on other aspects 
of the specific SUSY model possibly realized in nature.
Here the masses of the axino and the gravitino are of particular importance
since their thermal production can be very efficient in high $\TR$ scenarios
such as those explored in Fig.~\ref{Fig:extraradBBN}.

For example, in R-parity conserving scenarios 
in which the gravitino is the lightest SUSY particle (LSP),
the thermally produced gravitino density is limited from above 
by the dark matter density parameter 
$\omegaDM=\OmegaDM h^2\simeq 0.1$~\cite{Beringer}.
This disfavors $\TR\gtrsim 10^{10}\,\GeV$ and
translates into $\mgravitino\gtrsim 10~\GeV$
for $\TR=10^8\,\GeV$ and universal gaugino masses 
at the scale of grand unification of $m_{1/2}\sim 0.5~\TeV$;
cf.~Fig.~2 in Ref.~\cite{Pradler:2006hh}.
For $\msaxion\sim\mgravitino$, as expected in gravity-mediated SUSY breaking,
this cosmological constraint will then challenge  
the $\Delta\neff\sim 1$ explanation presented in Fig.~\ref{Fig:extraradBBN}
for $\TR=10^8\,\GeV$ and disfavor the one for $\TR=10^{10}\,\GeV$.
Depending on the next-to-lightest SUSY particle (NLSP),
even more restrictive upper limits on $\TR$ are possible; 
cf.~\cite{Steffen:2008qp} and references therein.
Additional limits related to axino cosmology can be evaded, e.g.,
for $\maxino\gtrsim 2~\TeV$, 
a gluino mass of $m_{\tilde{g}}\sim 1~\TeV$,
and $\fax\sim 10^{10}\,\GeV$.
Although possibly somewhat contrived from the model building point of view,
the heavy axinos then decay typically before the NLSP freeze-out
and the emitted sparticles will be thermalized such that 
the constraints associated with the NLSP will not be tightened.
If a sneutrino is the NLSP
(for which the NLSP-related constraints are rather 
mild~\cite{Feng:2004mt,Kanzaki:2006hm,Ellis:2008as}),
the shown $\Delta\neff\sim 1$ explanation for $\TR=10^8\,\GeV$ 
can thereby turn out to be viable for $\msaxion\sim\mgravitino\sim 10~\GeV$,
where cold dark matter can reside in thermally produced gravitinos.

In the alternative axino LSP case, one often finds more restrictive $\TR$ 
constraints imposed by the dark matter 
constraint~\cite{Covi:2001nw,Brandenburg:2004du,Strumia:2010aa,Freitas:2009fb}
and also additional $\fax$ constraints depending on the properties of the 
NLSP~\cite{Freitas:2009fb,Freitas:2009jb,Freitas:2011fx}.
Interestingly, these $\TR$ constraints can be avoided 
in the case of a light axino LSP with $\maxino<0.2~\keV$ 
(cf.\ Fig.~6 in~\cite{Brandenburg:2004du}).
Moreover, additional $\TR$ constraints from BBN-imposed limits 
on hadronic/electromagnetic energy injection from late decaying gravitinos 
can be evaded if the gravitino is the NLSP~\cite{Olive:1984bi,Asaka:2000ew}.
In such a setting, the lifetime of the gravitino NLSP is 
$\tau_{3/2}\sim 10^9\,\seconds\,(10^2\,\GeV/m_{3/2})^3$
and governed by its decay into the axino LSP 
and an axion~\cite{Olive:1984bi,Asaka:2000ew}.
While both of which are too weakly interacting to reprocess primordial nuclei,
the emitted particles can contribute to $\Delta\neff$ 
at cosmic times $t>\tau_{3/2}$~\cite{Ichikawa:2007jv}.
Upper limits on $\Delta\neff$ imposed by CMB + LSS studies 
have thereby been found to imply 
$\TR\lesssim 10^{11}\,\GeV$ at the $5\sigma$ level
for $m_{3/2}=100~\GeV$ and 
$m_{\tilde{g}}\sim 1~\TeV$~\cite{Hasenkamp:2011em}.
This limit can be overly conservative 
since it does not include $\Delta\neff$ from saxion decays into axions.
For $m_{\tilde{g}}\sim 1~\TeV$, 
which is still allowed by the ongoing LHC sparticle searches,
$m_{3/2}\sim 100~\GeV$, and $\TR\sim 10^{10}\,\GeV$,
gravitino decays into axions and axinos have been found to lead to
$\Delta\neff\sim 0.6$ 
but only at times well after the BBN 
epoch~\cite{Ichikawa:2007jv,Hasenkamp:2011em}. 
Taking into account the additional contribution to $\Delta\neff$
at such late times from saxion decays
(which we consider explicitly below), 
we find that this point in parameter space remains to be allowed.
This implies viability of the corresponding explanation 
of $\Delta\neff\sim 1$  at the onset of BBN 
for $\TR=10^{10}\,\GeV$ shown in Fig.~\ref{Fig:extraradBBN}.
Here one can easily accommodate also 
the small additional contribution of 
$\Delta\neff\lesssim 0.017$
provided by light thermal axinos
at the onset of BBN~\cite{Freitas:2011fx}.
For $\fax\sim 10^{12}\,\GeV$, 
cold dark matter can then reside in the form of an axion condensate
(cf.\ Fig.~\ref{Fig:Omegah2Axion} below) 
whereas axinos will be hot dark matter~\cite{Brandenburg:2004du}
with associated LSS constraints imposing 
$\maxino\lesssim 37~\eV$~\cite{Freitas:2011fx}.
While the lightest ordinary sparticle (LOSP) can still be long lived, 
BBN constraints related to its decay can be evaded.
For the stau LOSP case, this is illustrated explicitly 
in Fig.~21 of Ref.~\cite{Freitas:2011fx}.
Thereby, one arrives at viable scenarios with 
different $\Delta\neff$ predictions at the onset of BBN and much later,
in which even $\TR\sim 10^{10}\,\GeV$ is possible, e.g., 
allowing for the explanation of the baryon asymmetry 
via thermal leptogenesis~\cite{Buchmuller:2005eh}.

\subsection{CMB and LSS}
\label{Sec:CMBandLSS}

Axions from saxion decays can contribute to $\Delta\neff$ 
at the CMB decoupling epoch, even for $T_\saxion\gtrsim 1~\MeV$, 
as described below~\eqref{Eq:TNRaNTP}.
Extra radiation at that epoch delays the time of radiation-matter equality
and is probed by studies of the CMB anisotropies and the LSS distribution.
Here hints towards $\neff\gtrsim3$ have been found that are more pronounced
than those from BBN considered above; 
see~\cite{Hamann:2007pi,Reid:2009nq,Komatsu:2010fb,Hamann:2010pw,GonzalezGarcia:2010un} and references therein. 
For example, the Wilkinson Microwave Anisotropy Probe (WMAP) collaboration finds 
a 68\% credible interval of 
$\neff=4.34_{-0.88}^{+0.86}$~\cite{Komatsu:2010fb} 
when combining their 7-year data with measurements of 
the baryonic acoustic oscillation (BAO) scale 
and todays Hubble constant $H_0$. 
Another precision cosmology study arrives 
at a 95\% credible interval of 
$\neff=4.78_{-1.75}^{+1.86}$~\cite{Hamann:2010pw}
when combining CMB data 
with data from the Sloan Digital Sky Survey data-release 7 
halo power spectrum (HPS) and the Hubble Space Telescope (HST).
Based on this combined CMB + HPS + HST data set, 
we use the mean for $\Delta\neff$
and the 95\% CL upper limit on $\Delta\neff$,
as quoted in Table~\ref{Tab:Neffconstrains},
to explore implications for the considered SUSY axion models.

\begin{figure}[t]
\centering
\includegraphics[width=0.45\textwidth,clip=true]{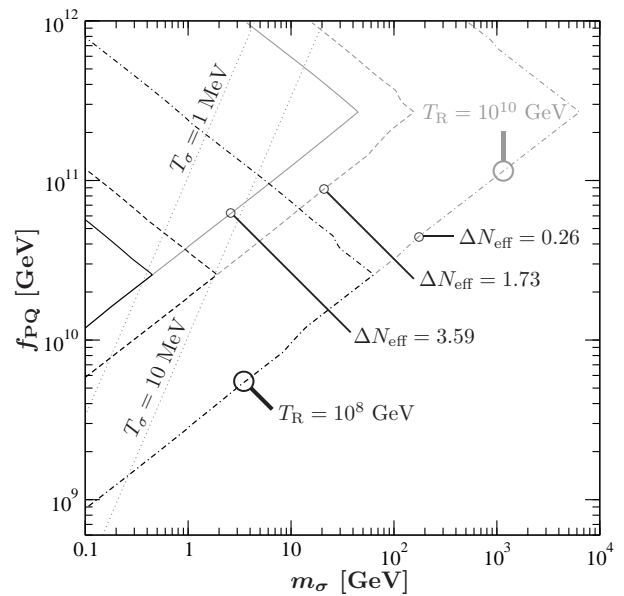}
\caption{{Contours of $\Delta\neff$ at $T\ll 1~\MeV$ 
provided by non-thermally produced axions from decays of
thermal saxions for $x=1$ and 
$\TR=10^8\,\GeV$ (black) and $10^{10}\,\GeV$ (gray).
The solid curve shows the $2\sigma$ limit $\Delta\neff=3.59$
and the dashed curve the mean $\Delta\neff=1.73$ 
based on the CMB + HPS + HST result~\cite{Hamann:2010pw}
quoted in Table~\ref{Tab:Neffconstrains}.
The dash-dotted curve indicates $\Delta\neff = 0.26$ 
which is the expected 68\% CL sensitivity 
of the Planck satellite mission~\cite{Perotto:2006rj,Hamann:2007sb}.
On the dotted lines, $T_\saxion=1$ and $10~\MeV$,
as in Fig.~\ref{Fig:extraradBBN}.}}
\label{Fig:extraradCMB}
\end{figure}

Evaluating $\Delta\neff(T)$ from \eqref{Eq:DeltaNeff1} 
and~\eqref{Eq:DeltaNeff2} 
for $x=1$ and at $T\ll 1~\MeV$,%
\footnote{Note that our theoretical results for $\Delta\neff(T)$ 
at $T\sim 1~\MeV$ and at $T\ll 1~\MeV$ agree. 
The $T$ dependence in~\eqref{Eq:DeltaNeff1} and~\eqref{Eq:DeltaNeff2}
results from the factor $(T/T_{\nu})^4[g_{*S}(T)/10.75]^{4/3}$.
This factor equals one for $T\sim 1~\MeV$, where $T_{\nu}=T$ and $g_{*S}(T)=10.75$,
and for $T\ll 1~\MeV$, where $T_{\nu}=(11/4)^{-1/3}T$ and $g_{*S}(T)=3.91$.}
we obtain the $\Delta\neff$ contours for $\tr=10^8\,(10^{10})\,\GeV$ 
as shown by the black (gray) lines in Fig.~\ref{Fig:extraradCMB}.
The solid lines indicate the upper limit $\Delta\neff=3.59$ 
with the $m_\saxion$--$\fax$ parameter regions 
to their left disfavored at the $2\sigma$ level by the CMB + HPS + HST data set.
The dashed lines show the corresponding mean $\Delta\neff=1.73$
and the dash-dotted lines $\Delta\neff=0.26$.
The latter is the expected 68\% CL accuracy 
of the Planck satellite mission~\cite{Perotto:2006rj,Hamann:2007sb}
mentioned already in the Introduction.
To guide the eye, we show again the $T_\saxion=1$ and $10~\MeV$ contours as dotted lines. 
Here we can provide the $\Delta\neff$ contours 
also in the region with $T_\saxion=1~\MeV$.
However, as described in the Introduction, 
additional restrictive constraints are expected in that region.

In comparison to the BBN-imposed limits,
one finds that the shown $2\sigma$ limit from precision cosmology
disfavors basically the same parameter region as the
conservative $3\sigma$ limit shown in Fig.~\ref{Fig:extraradBBN}.
The results from precision cosmology thereby 
allow for a slightly larger $\Delta\neff$ 
at given values of $\msaxion$, $\fax$, and $\TR$.
Indeed, comparing the mean and the $2\sigma$ limit 
in Table~\ref{Tab:Neffconstrains}
with the posterior maxima and the $3\sigma$ limits from the BBN study,
one finds a potential hint towards a $\Delta\neff$ value at $T\ll 1~\MeV$
that is higher than the one at $T\sim 1~\MeV$.
As discussed already in the preceding section, 
this may be a first hint for $\Delta\neff\sim 1$ already prior to BBN 
due to axions from saxion decays 
plus an additional late contribution 
from gravitino NLSP decays into axions and LSP axinos 
such that $\Delta\neff\sim 2$ at the time of CMB recombination.
With the expected $\Delta\neff$ sensitivity of the Planck satellite mission,
this possibility will be tested further soon.
Moreover, for scenarios in which
axions from saxion decays are the only significant source 
for $\Delta\neff$, the Planck results will allow us 
to probe significant regions of the $\msaxion$--$\fax$ parameter space
which have not been accessible by $\Delta\neff$ studies so far.
For fixed $\TR$ values, 
this is indicated by the dot-dashed lines in Fig.~\ref{Fig:extraradCMB}.
%

\begin{figure}[t]
\centering
\includegraphics[width=0.45\textwidth,clip=true]{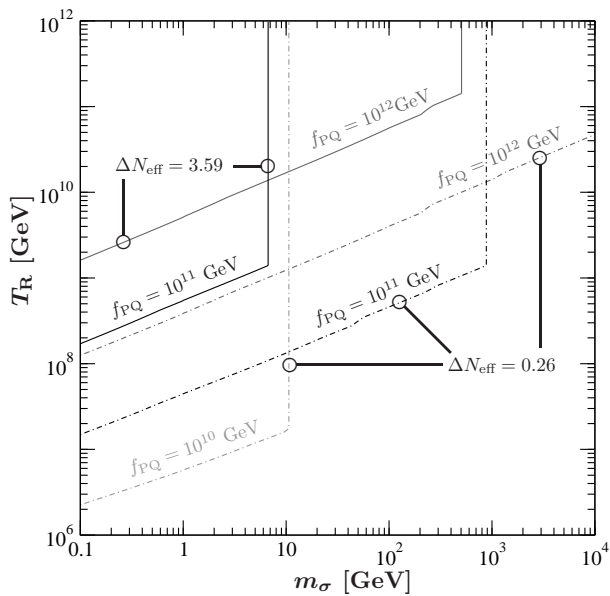}
\caption{{Contours of $\Delta\neff$ at $T\ll 1~\MeV$ 
provided by non-thermally produced axions from decays of
thermal saxions for $x=1$ and $\fax=10^{10}\,\GeV$ (light gray), 
$10^{11}\,\GeV$ (black), and $10^{12}\,\GeV$ (gray).
The solid lines show the $2\sigma$ CMB + HPS + HST constraint
$\Delta\neff<3.59$ 
which imposes upper limits on the reheating temperature $\TR$.
For $\fax=10^{10}\,\GeV$, the $\TR$ limit 
appears at smaller $\msaxion$
outside of the considered range.
The dash-dotted contours indicate the expected Planck sensitivity of 
$\Delta\neff=0.26$.}}
\label{Fig:TrLimits}
\end{figure}

The limits shown in Figs.~\ref{Fig:extraradBBN} and~\ref{Fig:extraradCMB}
in the $\msaxion$--$\fax$ plane for fixed $\TR$ values
can be translated into upper limits on the reheating temperature $\TR$.
In Fig.~\ref{Fig:TrLimits} the solid lines show the upper limits on $\TR$ 
imposed by the $2\sigma$ CMB + HPS + HST constraint 
$\Delta\neff<3.59$
as a function of $m_\saxion$ 
for $x=1$ 
and $\fax=10^{11}\,\GeV$ (black) and $10^{12}\,\GeV$ (gray).
%
%
The expected Planck sensitivity 
$\Delta\neff=0.26$
is indicated by the corresponding dash-dotted lines
and in light gray for $\fax=10^{10}\,\GeV$.
Note that the upper limit does not show up for the latter $\fax$ value
in the considered $\msaxion$ range.
The $\TR$ dependence of the contours is described 
by~\eqref{Eq:DeltaNeff2}
and disappears for cosmological scenarios with $\TR\gtrsim\TD$ 
where~\eqref{Eq:DeltaNeff1} applies.
Here we should stress that the shown upper limits on $\TR$
rely on non-thermally produced axions from decays of thermal saxions
providing the only significant contribution 
to $\Delta\neff$ at $T\ll 1~\MeV$.
In scenarios with additional sizable contributions 
(e.g., from late gravitino NLSP decays into an axino LSP and the axion),
the considered extra radiation constraint 
will impose more restrictive $\TR$ limits.
Nevertheless, the shown upper limits will remain to be applicable
as conservative guaranteed limits.

Let us compare our results shown in Fig.~\ref{Fig:TrLimits} 
with existing results.
For example, 
the yellow curve in Fig.~5(a) of Ref.~\cite{Kawasaki:2007mk} 
presents an upper $\TR$ limit imposed by $\Delta\neff\leq 1$
that disfavors basically $\TR>10^6\,\GeV$ for $\fax=10^{10}\,\GeV$
and the whole $\msaxion$ range considered above.
With the assumed initial saxion field displacement 
of $\sigma_\text{i}\sim\fax$,
that existing limit is governed also 
by thermal saxions that decay into axions.
However, we find that it is overly restrictive
due to the omission of the factor 
$[g_{*S}(T_\saxion)/g_{*S}(T)]^{4/3}$
in Eq.~(24) of Ref.~\cite{Kawasaki:2007mk}.%
\footnote{A similar comment can be found in Ref.~\cite{Hasenkamp:2011xh}
which refers to the same finding. We thank J.~Hasenkamp for clarification.}
Thereby, our $\Delta\neff$ expression~\eqref{Eq:DeltaNeffYield}
shows different dependences on $g_{*S}(T)$ and $g_{*S}(T_\saxion)$.
Remaining differences are due to the result 
from our explicit calculation 
of the thermal saxion production rate 
and the different definitions of the PQ scale 
addressed already in footnotes~2 and~5 above.
As a result, we find that a large part 
of the $(\msaxion,\,\fax,\,\TR)$ region
previously thought to be excluded is actually not restricted 
by the amount of additional radiation from late decays of thermal saxions.

\section{Relic axion density}
\label{Sec:AxionDensity}
%
We find it instructive to compare the density parameters
of three different axion populations that can be present today
in the considered SUSY axion models: 
(i)~$\Omega\ax^\text{eq/TP}h^2$ of thermal relic/thermally produced axions, 
(ii)~$\Omega\ax^\text{NTP}h^2$ of non-thermally produced axions 
from decays of thermal saxions, 
and (iii)~$\Omega\ax^\text{MIS}h^2$ of the axion condensate 
from the misalignment mechanism.
The latter originates from coherent oscillations 
of the axion field after it acquires a mass due to instanton effects 
at $T\lesssim 1~\GeV$.
This is the axion population that can provide 
the cold dark matter in our Universe,
as mentioned at the end of Sect.~\ref{Sec:BBN}.
For details on this misalignment mechanism 
we refer to~\cite{Sikivie:2006ni,Kim:2008hd,Beltran:2006sq} 
and references therein.
Here we quote the density parameter,
\be
\Omega\ax^\text{MIS}h^2
\sim 
0.15\,\theta_i^2 
\left(\frac{\fax}{10^{12}\,\GeV}\right)^{7/6},
\label{Eq:OmegaAxionMIS}
\ee
which is governed by 
the initial misalignment angle $\theta_i$ of the axion field.
This expression applies to non-SUSY and SUSY settings.
In the considered case in which the PQ symmetry breaks before inflation 
and is not restored thereafter, $\TR<\fax$, a single $\theta_i$ value 
will enter~\eqref{Eq:OmegaAxionMIS}.
The axion condensate cannot be thermalized 
by  processes such as those considered 
in Sect.~\ref{Sec:ThermalAxionProduction}
and the respective back reactions
since those processes proceed at negligible rates 
at $T\lesssim 1~\GeV$ for $\fax$ respecting~\eqref{Eq:fa_Limit}.%
\footnote{Recent studies explore the possibility that cold dark matter axions 
form a Bose-Einstein condensate~\cite{Sikivie:2009qn,Erken:2011dz,Erken:2011xj}. 
It is argued in these studies that the necessary condition of thermal equilibrium
can be established via gravitational axion self-interactions 
when $T$ reaches approximately $500~\eV(\fax/10^{12}\,\GeV)^{1/2}$.
This finding relies on the presence of a condensed regime at late times,
in which the transition rate between momentum states is large 
compared to their spread in energy.
Our study can neither reaffirm nor contradict this finding 
since our investigations are based on the usual Boltzmann equation
and thus restricted to the particle kinetic regime,
in which the opposite hierarchy holds.} 
Accordingly, $\Omega\ax^\text{MIS}h^2$ can coexist 
with $\Omega\ax^\text{eq/TP}h^2$ and $\Omega\ax^\text{NTP}h^2$, 
which we calculate in the following.

Since thermal relic and thermally produced axions 
have (basically) a thermal spectrum, 
one can describe the associated density parameter approximately by
\be
\Omega\ax^\text{eq/TP} h^2 
\simeq
\sqrt{\langle p_{\axion,0}^\text{th}\rangle^2+\maxion^2}\,\,
Y\ax^\text{eq/TP}\,\,
s(T_0) h^2/\rho_c ,
\label{Eq:OmegaAxionEqTP}
\ee
where $\rho_c/[s(T_0)h^2]=3.6~\eV$ and
$Y\ax^\text{eq/TP}=Y_\saxion^\text{eq/TP}$ 
as described in Sect.~\ref{Eq:SaxionAxionYield}. 
With the present CMB temperature of $T_0=0.235~\meV$
and an axion temperature today of
$T_{\axion,0}=[g_{*S}(T_0)/228.75]^{1/3}\,T_0\simeq 0.06~\meV$,
the average momentum of thermal axions today is given by
$\langle p_{\axion,0}^\text{th}\rangle=2.701\,T_{\axion,0}$.
When comparing this momentum with the axion mass $\maxion$, 
one finds that this axion population is still relativistic today 
for $\fax\gtrsim 10^{11}\,\GeV$.
At and before the CMB decoupling epoch,
$T\gtrsim 1~\eV$,
axions from thermal processes were relativistic 
for $\fax$ in the full allowed range~\eqref{Eq:fa_Limit}.
In the considered SUSY settings, 
they contribute at most 
\be
\Delta\neff(T) 
=
\frac{4}{7}\, 
\left(\frac{T_{\axion}}{T_\nu}\right)^4
=
\frac{4}{7}\, 
\left[\frac{g_{*S}(T)}{228.75}\right]^{4/3}
\left(\frac{T}{T_\nu}\right)^4 ,
\label{Eq:DeltaNeffAxionEq}
\ee
i.e., $\Delta\neff(T)\leq 0.0097$ for $10~\MeV\gtrsim T\gtrsim 1~\eV$,
which is far below the Planck sensitivity and 
easily accommodated by the $\Delta\neff$ limits discussed above.

The density parameter of non-thermal axions emitted in late decays 
of saxions from thermal processes reads
\be
\Omega\ax^\text{NTP}h^2 
= 
2\,
\sqrt{(p_{\axion,0}^\text{NTP})^2+\maxion^2}\,\,
Y_\saxion^\text{eq/TP}\,\, 
s(T_0)h^2/\rho_c
\label{Eq:OmegaAxionNTP}
\ee
with the present momentum of these axions given by
\be
p_{\axion,0}^\text{NTP} 
=  
\frac{\msaxion}{2} 
\left[\frac{g_{*S}(T_0)}{g_{*S}(T_\saxion)}\right]^{1/3} 
\frac{T_0}{T_\saxion},
 \ee
when applying the sudden decay approximation.
Thus, $\Omega\ax^\text{NTP}h^2$ will depend on $\msaxion$
if these axions are still relativistic today, i.e.,
when $T_0\gtrsim\TNR$ given by~\eqref{Eq:TNRaNTP} above.
As extensively discussed in the previous section, 
this non-thermal axion population can provide a
significant contribution to $\Delta\neff$
prior to BBN and thereafter.
In fact, 
one can use~\eqref{Eq:DeltaNeffAxionNTP} 
to relate $\Omega\ax^\text{NTP}h^2$ to this $\Delta\neff$: 
\begin{eqnarray}
&&\Omega\ax^\text{NTP}h^2 
= 
\Bigg\{ 
1.1\times 10^{-11}
\left( \frac{10^{9}\,\GeV}{\fax} \right)^{\!2}
\left( \frac{Y_{\saxion}^\text{eq/TP}}{10^{-3}} \right)^{\!2}  
\nonumber \\
& &
+\,\, 
3.2\times 10^{-11}  
\left(\frac{T_\nu}{T}\right)^{\!8}
\left[\frac{10.75}{g_{*S}(T)}\right]^{8/3} \!\!\!
\Delta\neff^2(T)
\Bigg\}^{\!1/2}\!\!\!\!,
\label{Eq:OmegaNeff}
\end{eqnarray}
where the $\msaxion$ dependence is now absorbed into $\Delta\neff(T)$.
Thus, the discussed $\Delta\neff$ constraints translate directly
into upper limits on $\Omega\ax^\text{NTP}h^2$.
For $T\lesssim 1~\MeV$ and $\fax$ 
such that the first term on the right-hand side of~\eqref{Eq:OmegaNeff} 
is negligible,
those limits are described by 
$\Omega\ax^\text{NTP}h^2=5.7\times 10^{-6}\,\Delta\neff$.
(This applies to thermal axions as well
if they are still relativistic today.)

\begin{figure}[t]
\centering
\includegraphics[width=0.45\textwidth,clip=true]{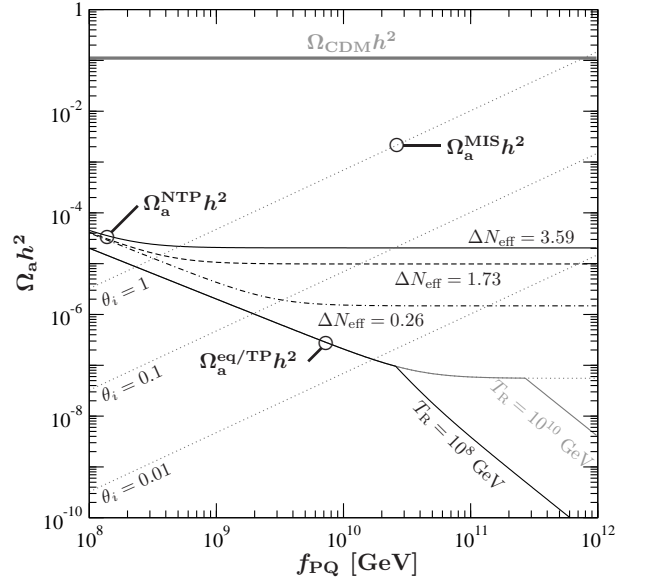}
\caption{{The density parameters
of the axion condensate from the misalignment mechanism 
$\Omega\ax^\text{MIS}h^2$ 
for $\theta_i=0.01$, 0.1, and 1
(dotted lines),  
of non-thermally produced axions from decays of thermal saxions 
$\Omega\ax^\text{NTP}h^2$ 
for $\Delta\neff=3.59$ (solid), 1.73 (dashed), and 0.26 (dash-dotted),
and of thermal relic/thermally produced axions 
$\Omega\ax^\text{eq/TP}h^2$ 
for $\TR=10^8$ (black) and $10^{10}\,\GeV$ (gray).
The dotted line connected to the latter 
indicates $\Omega\ax^\text{eq}h^2$
for larger $\TR$ with $\TR>\TD$.
The dark matter density parameter
$\Omega_\text{CDM}h^2\simeq0.1$~\cite{Beringer}
is indicated by the horizontal gray bar.
As in Fig.~\ref{Fig:extraradCMB},
we consider $\Delta\neff$ at $T\ll 1$ and show values based on 
the CMB + HPS + HST result~\cite{Hamann:2010pw} 
quoted in Table~\ref{Tab:Neffconstrains} 
and the expected 68\%~CL sensitivity of the Planck satellite mission~\cite{Perotto:2006rj,Hamann:2007sb}.
}}
\label{Fig:Omegah2Axion}
\end{figure}

Figure~\ref{Fig:Omegah2Axion} 
%
%
shows $\Omega\ax^\text{NTP}h^2$ contours that correspond to 
$\Delta\neff$ values 
of 3.59 (solid), 1.73 (dashed), and 0.26 (dash-dotted).
As in Fig.~\ref{Fig:extraradCMB}, 
these values are motivated by 
the CMB + HPS + HST result~\cite{Hamann:2010pw} 
quoted in Table~\ref{Tab:Neffconstrains} 
and the expected 68\% CL sensitivity 
of the Planck satellite mission~\cite{Perotto:2006rj,Hamann:2007sb}.
Contours of $\Omega\ax^\text{eq/TP}h^2$
are shown for $\TR=10^8$~$(10^{10})$~GeV 
by the solid black (gray) lines
and for larger $\TR>\TD$ by the unlabeled dotted line.
On this dotted line at $\fax>10^{11}\,\GeV$, 
\eqref{Eq:DeltaNeffAxionEq} applies so that $\Delta\neff=0.0097$ 
and $\Omega\ax^\text{eq}h^2=5.5\times 10^{-8}$
reside in thermal relic axions
that are still relativistic today.
The labeled dotted lines  
indicate $\Omega\ax^\text{MIS}h^2$ 
of the axion condensate from the misalignment mechanism 
for $\theta_i=0.01$, 0.1, and 1.
For $\theta_i\sim1$ and $\fax\sim 10^{12}\,\GeV$, 
this cold axion population can explain the dark matter density
$\Omega_\text{CDM}h^2\simeq 0.1$~\cite{Beringer}
displayed by the gray bar.

Considering the $2\sigma$ limit $\Delta\neff< 3.59$ 
in Fig.~\ref{Fig:Omegah2Axion}, 
one sees that it constrains $\Omega\ax^\text{NTP}h^2$ 
to values that stay below the photon density 
$\Omega_{\gamma}h^2\simeq 2.5\times 10^{-5}$~\cite{Beringer}.
Remarkably, Planck results are expected 
to probe even much smaller $\Omega\ax^\text{NTP}$.
The testable values can be
as small as an order of magnitude below $\Omega_{\gamma}$
if axions emitted in decays of thermal saxions are 
the only significant contribution to $\Delta\neff$.
In contrast and similarly to the non-SUSY case~\cite{Graf:2010tv},
it will remain to be extremely challenging to probe
the axion population from thermal processes 
with its small contribution of $\Delta\neff\lesssim 0.01$.

Note that $\msaxion$ changes along the $\Omega\ax^\text{NTP}h^2$ 
curves in Fig.~\ref{Fig:Omegah2Axion} 
for fixed $\TR$ and $x$
since we indicate results for fixed values of $\Delta\neff$.
Indeed, additional BBN constraints can disfavor parts of the 
shown contours when $T_\sigma<1~\MeV$.
For $T_\sigma>1~\MeV$, BBN constraints on $\Delta\neff$ --
such as the ones considered in Fig.~\ref{Fig:extraradBBN} --
can also be displayed in terms of $\Omega\ax^\text{NTP}h^2$.
On the logarithmic scale considered in Fig.~\ref{Fig:Omegah2Axion}, 
they are similar to the shown ones. 

Taking into account the relation between $\fax$ and $m_a$, 
the analog of a Lee--Weinberg curve is given by
$\Omega\ax h^2
\geq
\Omega\ax^\text{MIS}h^2
+\Omega\ax^\text{NTP}h^2
+\Omega\ax^\text{eq/TP}h^2$
and can be inferred from Fig.~\ref{Fig:Omegah2Axion}.
Depending on the initial displacement 
of the saxion field from the vacuum, $\sigma_i$,
and on the mass spectrum,
there can be additional contributions to 
the axion density parameter, e.g.,
from decays of the saxion condensate into axions
and/or a gravitino NLSP into axions and LSP axinos.
In such cases, sizable additional
contributions also to $\Delta\neff$ are possible
which will affect
the $\Omega\ax^\text{NTP}h^2$ contours 
in Fig.~\ref{Fig:Omegah2Axion}. 
Thus the shown contours should be understood
as conservative maximum values.

One can consider Fig.~\ref{Fig:Omegah2Axion} as a SUSY generalization
of Fig.~4 in Ref.~\cite{Graf:2010tv}, which allows one
to infer the axion analog of the Lee--Weinberg curve in non-SUSY scenarios.
Whereas $\Omega\ax^\text{eq/TP}h^2$ can govern the axion density
for small $\theta_i$ and/or small $\fax$ in non-SUSY scenarios~\cite{Graf:2010tv},
we find $\Omega\ax^\text{NTP}h^2\gtrsim 2\,\Omega\ax^\text{eq/TP}h^2$
in the considered SUSY scenarios.
This can be seen in Fig.~\ref{Fig:Omegah2Axion} 
and when comparing~\eqref{Eq:OmegaAxionEqTP} 
and~\eqref{Eq:OmegaAxionNTP}.
If SUSY and a hadronic axion model are realized in nature, 
the axion density parameter can thus be governed by
non-thermal axions from decays of thermal saxions
and/or the axion condensate from the misalignment mechanism.
Interestingly, both of these populations may be accessible experimentally:
While signals of the axion condensate are expected
in direct axion dark matter searches~\cite{Carosi:2007uc}, 
the findings of $\Delta\neff$ studies may already be first hints
for the existence of non-thermal axions from saxion decays.

\section{Conclusion}
\label{Sec:Conclusion}

We have explored 
thermal production processes of axions and saxions
in the primordial plasma,
resulting axion populations and their manifestations 
in the form of extra radiation $\Delta\neff$ 
prior to BBN and well thereafter.
The considered SUSY axion models are attractive for a number of reasons.
For example, they allow 
for simultaneous solutions of the strong CP problem,
the hierarchy problem, and the dark matter problem.

Here we have focussed on the saxion, which can be a late decaying particle
and as such be subject to various cosmological constraints.
We find that the saxion decay into two axions is often the dominating one.
For a saxion mass of $\msaxion\gtrsim 1~\GeV$, 
such decays occur typically before the onset of BBN.
We have shown that the emitted axions can then still be relativistic
at the large scattering surface.
Thereby, they can provide sizable contributions 
to extra radiation $\Delta\neff$ that is testable
in BBN studies 
and in precision cosmology of the CMB and the LSS.

We have aimed at a consistent description of both 
the thermal axion/saxion production 
and of saxion decays into axions.
This has motivated our careful derivations of the Lagrangian
$\mathcal{L}_\text{\PQ}^\text{int}$
that describes the interactions of the 
PQ multiplet with quarks, gluons, squarks, and gluinos
and of  
$\mathcal{L}_\text{\PQ}^\text{kin}$
that describes 
the interactions of saxions with axions 
in addition to their kinetic terms.
The requirement of canonically normalized kinetic terms
defines the scale $\vax$, which governs 
the saxion-axion-interaction strength
together with another PQ-model-dependent parameter $x\lesssim 1$.
On the other hand,
the form of the effective axion-gluon-interaction term
defines the PQ scale $\fax$.
Considering the emergence of this term
from loops of heavy KSVZ fields
in an explicit hadronic axion model, 
we find $\fax=\sqrt{2}\vax$.
This is in contrast to numerous existing studies 
which treat $\vax$ and $\fax$ synonymously.

Relying on the derived form of $\mathcal{L}_\text{\PQ}^\text{int}$,
we have calculated the thermal production rates of saxions and axions 
and the resulting yields in the hot early Universe.
Despite differences in the interaction terms,
we find that the rate for thermal saxion production
agrees with the one for thermal axion production.
This implies an agreement also of 
the calculated thermally produced yields
and of our estimates of the decoupling temperatures $\TD$.
By applying HTL resummation~\cite{Braaten:1989mz} 
and the Braaten--Yuan prescription~\cite{Braaten:1991dd},
finite results are obtained in a gauge-invariant way 
consistent to leading order in the coupling constant 
and screening effects are treated
systematically.

Using our result for the thermally produced saxion yield,
we have calculated $\Delta\neff$ provided in the form
of axions from decays of thermal saxions.
This has allowed us to demonstrate 
that such a  $\Delta\neff$ contribution
can indeed explain the trends towards extra radiation beyond the SM 
seen in recent studies of BBN, CMB, and LSS.

To account for the current PDG recommendation 
for the free neutron lifetime, 
$\tau_{\mathrm{n}}=880.1\pm1.1~\seconds$~\cite{Beringer},
we have performed a BBN likelihood analysis
with {\tt PArthENoPE}~\cite{Pisanti:2007hk} 
and
based on recent insights on the primordial abundances of 
$^4$He~\cite{Izotov:2010ca,Aver:2010wq} 
and D~\cite{MNR:MNR13921}.
For $\Delta\neff$ at the onset of BBN,
we thereby obtain 
posterior maxima of 0.76 and 0.77
and $3\sigma$ upper limits of 1.97 and 3.53
with the $Y_\text{p}$ results
of~\cite{Izotov:2010ca} and~\cite{Aver:2010wq}, respectively.
When comparing these values with results 
from studies of the CMB and LSS, we find
that the latter provide compatible but
more pronounced hints for extra radiation.
For example, the precision cosmology study 
of~\cite{Hamann:2010pw}
reports a mean of 1.73 and a $2\sigma$ limit 3.59
for $\Delta\neff$ at $T\ll 1~\MeV$
when using the CMB + HPS + HST data set.

We have translated the upper limits on $\Delta\neff$ quoted above 
into bounds on $\fax$, $\msaxion$, and $\TR$.
These bounds can disfavor significant regions of the 
$\msaxion$--$\fax$ parameter plane in high $\TR$ scenarios.
However, we find that our limits leave open a considerable parameter region 
previously thought to be excluded~\cite{Kawasaki:2007mk}.
Significant parts of the allowed parameter region
have been identified, which will become accessible very soon 
with the upcoming results from the Planck satellite mission. 

The explanation of the above hints for extra radiation 
via axions from decays of thermal saxions
requires a relatively high reheating temperature
of $\TR\gtrsim 10^7~\GeV$ for $\msaxion\gtrsim 0.1~\GeV$.
Such high $\TR$ scenarios can be in conflict with cosmological constraints due to overly efficient thermal production of axinos and gravitinos.
To illustrate the viability of $\Delta\neff\sim 1$ from saxion decays,
we have described two exemplary SUSY scenarios
which allow for $\TR=10^8$ and $10^{10}~\GeV$:
\begin{itemize}
\item[(i)] With the gravitino LSP as cold dark matter and a sneutrino NLSP,
the presented $\Delta\neff\sim 1$ explanation for $\TR=10^8~\GeV$ 
can be viable for $\msaxion\sim\mgravitino\sim 10~\GeV$
and $m_{\tilde{g}}\sim 1~\TeV$.
This explanation requires $\fax\sim 10^{10}\,\GeV$
and heavy axinos, $\maxino\gtrsim 2~\TeV$, 
which decay prior to NLSP decoupling.
Here $\Delta\neff\sim 1$ is already present at the onset
of BBN and does not change thereafter.
Accordingly, we expect that the Planck results will
point to a $\Delta\neff$ value that is consistent with
the one inferred from BBN studies.
\item[(ii)] With a very light axino LSP, $\maxino\lesssim 37~\eV$, 
 as hot dark matter and a gravitino NLSP,
the $\Delta\neff\sim 1$ explanation for $\TR=10^{10}~\GeV$ 
can be viable for $\msaxion\sim\mgravitino\sim 100~\GeV$
and $m_{\tilde{g}}\sim 1~\TeV$.
Here this explanation requires $\fax\sim 10^{12}\,\GeV$
so that cold dark matter can be provided by
the axion misalignment mechanism.
With the stau as the LOSP, further potential
BBN constraints can be evaded.
Note that $\TR=10^{10}~\GeV$ allows 
for successful thermal leptogenesis.
The saxion decays give $\Delta\neff\sim 1$
already at the onset of BBN.
However, late gravitino NLSP decays 
into the axion and the axino LSP 
can provide an additional contribution
of $\Delta\neff\sim 1$ 
well after BBN~\cite{Ichikawa:2007jv,Hasenkamp:2011em}.
Thus, it will be interesting to see 
whether the Planck results confirm the trend towards
an excess of extra radiation 
that is more pronounced at late times.
For example, the finding of $\Delta\neff\sim 2$
at late times
will be a possible signature expected in this setting.
\end{itemize}

If a SUSY hadronic axion model is realized in nature, 
three different axion populations will be present today: 
thermally produced/thermal relic axions,
non-thermally produced axions from decays of thermal saxions,
and the axion condensate from the misalignment mechanism.
We have calculated and compared the associated density parameters.
The results allow us to infer the axion analog of the Lee--Weinberg curve.
For $\fax\gtrsim 10^{12}~\GeV$ and 
an initial misalignment angle of $\theta_i\sim 1$,
the axion density parameter is governed
by the axion condensate.
In that parameter region this population
may be accessible in direct axion dark matter searches.
For smaller $\fax$ and smaller $\theta_i$, 
axions from saxion decays can dominate 
the axion density parameter.
While it will be extremely challenging to probe thermal axions,
Planck may confirm $\Delta\neff$ signals of
this non-thermally produced population 
in the full allowed $\fax$ range.
Since the considered axion populations can coexist,
there is the exciting chance to see signals of both
axion dark matter and axion dark radiation 
in current and future experiments.

\section*{Acknowledgments}
%
We are grateful to S.~Halter, G.~Raffelt, S.~Sarikas, 
and Y.Y.Y.~Wong for valuable discussions.
This research was partially supported by the 
Cluster of Excellence ``Origin and Structure of the Universe.''


\end{document}